\documentclass{preprint}

\date{June 1st, 2006}

\usepackage{graphics}
\usepackage{amsfonts}
\usepackage{amsmath}
\usepackage{amssymb}

\usepackage{times}
\usepackage{mhequ}
\usepackage{mhenvs}

\newtheorem{example}[lemma]{Example}

\newcommand{\eqdef}{\stackrel{\mbox{\tiny def}}{=}}
\newcommand{\one}{\mathbf{1}}

\newcommand{\Eg}{\mathcal{E}}

\newcommand{\CC}{\mathbb{C}}

\newcommand{\EE}{\mathbb{E}}

\newcommand{\E}{\mathbf{E}}

\newcommand{\NN}{\mathbb{N}}
\newcommand{\RR}{\mathbb{R}}
\newcommand{\arctanh}{\text{arctanh}}

\renewcommand{\d}{\partial}
\newcommand{\dt}{\frac{d\ }{dt}}

\renewcommand{\theenumi}{\roman{enumi}}
\renewcommand{\theenumiii}{\arabic{enumiii}}

\begin{document}
\title{Simple Systems with Anomalous Dissipation\\ and Energy Cascade}

\author{Jonathan C. Mattingly\inst{1},  Toufic Suidan\inst{2}, and\\  Eric Vanden-Eijnden\inst{3}}

\institute{Department of Mathematics and CNCS, Duke University,
  Durham, NC 27708,
  USA. 
  \email{jonm@math.duke.edu} \and Mathematics Department, University
  of California, Santa Cruz, CA 95064, USA. 
  \email{tsuidan@ucsc.edu} \and Courant Institute,
  New York University, New York, NY 10012, USA. 
  \email{eve2@cims.nyu.edu} }

\maketitle
\begin{abstract}
  We analyze a class of dynamical systems of the type
  \begin{equation*}
      \dot a_n(t) = c_{n-1} a_{n-1}(t) - c_n a_{n+1}(t) + f_n(t), 
      \quad n \in \NN,\ a_0=0,
  \end{equation*}
  where $f_n(t)$ is a forcing term with $f_n(t)\not = 0$ only for
  $n\le n_\star<\infty$ and the coupling coefficients $c_n$ satisfy a
  condition ensuring the formal conservation of energy~$\frac12\sum_n
  |a_n(t)|^2$. Despite being formally conservative, we show that these
  dynamical systems support dissipative solutions (suitably defined)
  and, as a result, may admit unique (statistical) steady states when
  the forcing term $f_n(t)$ is nonzero. This claim is demonstrated via
  the complete characterization of the solutions of the system above
  for specific choices of the coupling coefficients~$c_n$. The
  mechanism of anomalous dissipations is shown to arise via a cascade
  of the energy towards the modes with higher~$n$; this is responsible
  for solutions with interesting energy spectra, namely $\EE |a_n|^2$
  scales as $n^{-\alpha}$ as $n\to\infty$. Here the exponents $\alpha$
  depend on the coupling coefficients $c_n$ and $\EE$ denotes
  expectation with respect to the equilibrium measure. This is
  reminiscent of the conjectured properties of the solutions of the
  Navier-Stokes equations in the inviscid limit and their accepted
  relationship with fully developed turbulence. Hence, these simple
  models illustrate some of the heuristic ideas that have been
  advanced to characterize turbulence, similar in that respect to the
  random passive scalar or random Burgers equation, but even simpler
  and fully solvable.
\end{abstract}

\section{Introduction and main results: Life starts after blow-up}
\label{sec:intro}\label{sec:Mainstory}

So little is understood about hydrodynamic turbulence that there is
not even consensus on what it is. However, most physicists would agree
on the following heuristic picture which has emerged from the works
Kolmogorov, Onsager, Richardson, etc~\cite{Frisch95}.  In this
picture, (fully developed) turbulence refers to the idealized state of
an incompressible fluid described by the Navier-Stokes equations in the
limit of vanishing molecular viscosity. In this limit, the
Navier-Stokes equations formally reduces to the Euler equations, and
the turbulent solutions should be the most regular solutions of the
Euler equations which dissipate energy.  This is referred to as
anomalous dissipation and is best visualized in the Fourier
representation.  There it corresponds to a cascade of energy from the
small wavenumbers (large spatial scales) where energy is injected
(either via the initial condition or by a forcing term in the
equation) towards larger and larger wavenumbers (smaller and smaller
scales), up to infinity where energy should eventually be dissipated.
It is also believed that the cascade of energy implies that the energy
spectrum of the turbulent solutions have a power law decay in the
wavenumber whose rate can be deduced by dimensional analysis and is
$\frac53$ in three-dimension of space.

Turbulence theory (as we shall refer to the heuristic picture above)
also discusses more advanced and more controversial topics such as
intermittency.  But, without even going into those, most
mathematicians would agree that a rigorous confirmation of the basic
predictions of turbulence theory is already a tremendous challenge.
The best known results on the Navier-Stokes and Euler equations which
corroborate the above were obtained
in~\cite{ConstantinETiti94OCE,Eyink01DTE,DuchonRobert00IED}. These
works only indicate that turbulence theory is not blatantly
inconsistent.  Simpler models, such as randomly forced Burgers
equation or Kraichnan's model of passive scalar advection (see e.g.
\cite{E01SH,FalkovichGaweddzkiVergassola01PFF} for reviews), have also
been used to demonstrate that parts of turbulence theory make sense in
terms of anomalous dissipation of the weak solutions of the inviscid
Burgers equation and the spectrum of energy of the solutions that this
implies. Even these simple models remain surprisingly complicated to
analyze and a full characterization of the statistical properties of
their solutions is still lacking.

One of the purposes of the present paper is to illustrate turbulence
theory on even simpler models.  Many (if not most) of the realistic
features have been neglected in our models. Yet, the models possess a
rich range of behaviors which depend on the details of the
interactions. They provide a simple class of exactly solvable models
which can be useful in understanding the inner workings of some energy
transfer mechanisms. The solutions of these models are also consistent
with much of the claims of turbulence theory. In a way, they offer a
setting for the skeptical mathematician to understand the motivation
behind these claims, and if this paper succeed in doing this, we will
have achieved our main goal.

Next, we introduce the models that we will investigate and we
summarize the principal results of the paper.  As we will see, the
most interesting and meaningful solutions of these models are
solutions which have blown-up, such that they have become
infinite in some norm. This justifies our claim that ``life starts
after blow-up'': Disregarding these solutions as nonsensical, as one
may be tempted to do at first sight, would, in fact, completely miss the
most interesting phenomena displayed by the models.

\subsection{A linear shell model}
\label{sec:B_Example}

Consider the equation
\begin{equation*}
  \dot a_n(t) = c\big[ (n-1)a_{n-1}(t) - n a_{n+1}(t)\big]
\end{equation*}
for $n\in \NN$ with the boundary condition $a_0(t)=0$ for all $t$.  If
$c>0$, we can rescale time to fix $c=1$; observe also that if $a_n$
satisfies the equations with the parameter $c< 0$ then $\hat a_n(t) =
(-1)^{n+1} a_n(t)$ satisfies the equations with parameter $|c|$.

In light of these considerations, we set $c=1$ and focus our
attention on
\begin{equation}
  \label{eq:B}
  \dot a_n(t) = (n-1)a_{n-1}(t) - n a_{n+1}(t) 
\end{equation}
for $n\in\NN$ with the boundary condition $a_0(t)=0$ for all $t$.

Although we will
see that this calculation is not always correct, on the formal level
one has that
\begin{equation}
  \label{eq:nodiss}
  \begin{aligned}
    \frac12\dt \sum_{n=1}^\infty |a_n(t)|^2&= \sum_{n=1}^\infty
    [(n-1)a_{n-1}(t) a_n(t) - na_{n+1}(t) a_n(t)]\\
    &=\sum_{n=1}^\infty n a_{n}(t)a_{n+1}(t) - \sum_{n=1}^\infty n
    a_n(t)a_{n+1}(t)=0 \; .
  \end{aligned}
\end{equation}
The second equality is only formal as it assumes that the sum
$\sum_{n=1}^\infty n a_n(t)a_{n+1}(t)$ is finite and absolutely
convergent. To understand this further, consider the evolution of the
partial sum $\sum_{n\leq N} |a_n(t)|^2$. For $N\in\NN$,
\begin{equation}\label{eq:dissipationModelOne}
  \frac12\dt \sum_{n \le N} |a_n(t)|^2= - N a_N(t)a_{N+1}(t).
\end{equation}
The validity of~\eqref{eq:nodiss} necessitates
\begin{equation}\label{eq:yesConserve}
  \lim_{N\to\infty} N a_N(t)a_{N+1}(t) =0.
\end{equation}
If this condition is not satisfied, then the formal manipulation
in~(\ref{eq:nodiss}) does \textit{not} hold and the seemingly
conservative coupling term in~(\ref{eq:B}) may become a source of
anomalous dissipation. We make the concept of anomalous dissipation
precise in Section~\ref{sec:anomalousdiss}. But, roughly speaking, it
is when seemingly conservative terms have a dissipative effect on the
system.

In the context of equation \eqref{eq:B}, anomalous dissipation seems
to require that the limit as $N \rightarrow \infty$ of the right hand
side of \eqref{eq:dissipationModelOne} be negative. In other words,
equation \eqref{eq:B} is dissipative at time $t$ if
\begin{equation}
  \label{eq:dissipativeDef}
  \liminf_{n \rightarrow \infty}  na_{n}(t)a_{n+1}(t) > 0.
\end{equation}
If we make the reasonable assumption that $\lim_n a_{n+1}/a_n \in
(0, \infty)$, then from~(\ref{eq:dissipativeDef}) the solution of
equation \eqref{eq:B} will be dissipative with a finite dissipation
rate provided that
\begin{equation}\label{eq:huristicAsymptotics}
  a_{n}(t) \eqsim  \frac1{\sqrt{n}} \quad \text{ as $n\to\infty$}.
  \footnote{%
    We say that $f(m) \eqsim g(m)$ as $m \rightarrow \infty$ if
    their exists an $m_1$ and $c \geq 1$
    so that if $m>m_1$ then $ \frac1c g(m) \leq  f(m)  \leq  c g(m)$.
    Similarly, we say that $f(m) \sim g(m)$ as $m \rightarrow m_0$ 
    if $\lim  f(m)/g(m) =1$ as $m\rightarrow m_0$. }
\end{equation} 
At this point some readers may be skeptical since one typically
considers equations like \eqref{eq:B} with initial data in $\ell^2$,
the space of square-summable sequences. However, we will see (Theorem
\ref{thm:simpleExistence} in Section \ref{sec:examplesFirst}) that
equation \eqref{eq:B} has solutions which exist for all time provided
\begin{equation*}
  \limsup_{n \rightarrow \infty} |a_n(0)|^{1/n}\leq 1\;.
\end{equation*}
This condition admits a large class of initial conditions
including those which scale as \eqref{eq:huristicAsymptotics}. In
Section \ref{sec:examplesFirst}, we will also see that \eqref{eq:B}
possesses a wide verity of behavior including conservative,
dissipative, and explosive solutions.

It might be tempting to dismiss these non-conservative solutions as
non-physical solution arising from pathological data. We now discuss
why is not the case.

Consider equation \eqref{eq:B} with a white-noise forcing  in
the first coordinate:
\begin{equation}
  \label{eq:Bforce}
  \dot a_n(t) = (n-1)a_{n-1}(t) - n a_{n+1}(t) + \one_{n=1} \dot W(t).
\end{equation}
where $\one_{n=m}$ is $1$ if $n=m$ and $0$ otherwise, and $W(t)$
denotes a standard Brownian motion, i.e. Gaussian process with mean zero and
covariance $\EE W(t)W(s) =\min(t,s)$.  If one were to accept the
formal calculations in \eqref{eq:nodiss}, showing energy conservation,
then
\begin{equation*}
  \EE \sum_{n=1}^\infty |a_n(t)|^2 =   \EE \sum_{n=1}^\infty
  |a_n(0)|^2 + t
\end{equation*}
if the energy is initially finite. Hence, in the forced system energy
seems to grow linearly with time and at $t=\infty$ one expects the
system to have infinite energy. These solutions which ``blow-up'' (in
the sense that they have infinite energy) are the most interesting and
relevant. In light of the discussion above, one might expect that the
energy of the system would grow to be infinite and arrange the $a_n$
so that the calculation in \eqref{eq:nodiss} is not valid since the
sum is not rearrangeable. Onsager would then predict that the system
would evolve to the state in which the $a_n$ decayed as fast as
possible but sill dissipated energy in the sense that
\eqref{eq:dissipativeDef} holds. The reasoning which leads to
\eqref{eq:huristicAsymptotics} strongly suggests that the $|a_n|$
should scale as $1/\sqrt{n}$. In fact, if the system is to reach some
equilibrium the effect of the dissipation must exactly balance that of
the forcing. Specifically, in the stochastic setting when the forcing
is $\dot W(t)\one_{n=1}$, $\lim_n n\, \EE (a_n(t)a_{n+1}(t))
\rightarrow 1$ as $t \rightarrow \infty$.

All of these conclusion turn out to be correct. In particular, in
Section \ref{sec:BForced} we prove that if $\sum_n |a_n(0)|^2 <
\infty$ then the solutions converge to a unique random variable
$a^{**}=(a_1^{**},a_2^{**},\cdots)$ which is Gaussian with mean zero
and whose distribution is the unique stationary measure for the
system. Furthermore, this equilibrium state has a structure which is consistent with the anticipated $1/\sqrt{n}$
scaling:
\begin{equation}
  \lim_{t \rightarrow \infty } \EE\,a_n(t)a_{m}(t)=
  \EE\,a_n^{**}a_{m}^{**}=\frac{1}{n+m-1}\ .
\end{equation}
We also show that similar behaviors are observed with a different type
of forcing. In particular, if the forcing is constant, 
\begin{equation}
  \label{eq:cstforce}
  \dot a_n(t) = (n-1)a_{n-1}(t) - n a_{n+1}(t) + \one_{n=1},
\end{equation}
then the system evolves to a unique steady state $a_n^*$ which scales
as
\begin{equation}
  \lim_{t \rightarrow \infty } a_n(t)=a_n^*= \frac{\sqrt{\pi}
    \Gamma(\frac{n}{2})}{n \Gamma(\frac{n-1}{2})}\sim
  \sqrt{\frac{\pi}{n}}\ .
\end{equation}
In summary, we see that the forced systems' energy grows linearly with
time if the energy is initially finite. Asymptotically, the system
rearranges itself so that it reaches a state which dissipates energy
at $t=\infty$.  This state is chosen so that the dissipation rate
matches the energy flux into the system from the forcing. Since the
energy flux is finite, this leaves $|a_n| \eqsim 1/\sqrt{n}$ as the
only choice.  A slower decay rate would produce an infinite rate of
dissipation and a faster decay rate would produce a system which
conserved energy since the limit in \eqref{eq:dissipativeDef} would be
zero.

\subsection{A second linear shell model}
\label{sec:secondModelIntro}

We introduce a second model which also exhibits interesting
but different ``blow-up'' behavior.  Consider
\begin{equation}
  \label{eq:C}
  \dot b_n = (n-1)(n-\tfrac12) b_{n-1} - n (n+\tfrac12) b_{n+1} 
\end{equation}
for $n\in\NN$ with the boundary condition $b_0(t)=0$ for all $t$.  As
in the previous subsection, the unforced equation \textit{formally}
conserves energy since
 \begin{equation}\label{eq:bNoDiss}
    \frac12\dt \sum_{n=1}^\infty b^2_n(t)= \sum_{n=1}^\infty
    [(n-1)(n-\tfrac12)b_{n-1}(t) b_n(t) - n(n+\tfrac12)b_{n+1}(t) b_n(t)]=0.
  \end{equation}
  This equality (as in the case of equation~\eqref{eq:B}) is only
  formal since, in general, the sum cannot be rearranged. As before, to
  gain insight we consider the partial sums. For $N\in\NN$,
\begin{align*}
  \frac12\dt \sum_{n \le N} b^2_n(t)&= - N(N+\tfrac12) b_N(t)b_{N+1}(t).
\end{align*}
With this in mind, (\ref{eq:C}) will be called conservative if
\begin{equation*}
  \lim_{N\to\infty}  N(N+\tfrac12) b_N(t)b_{N+1}(t) =0,
\end{equation*}
and dissipative if
\begin{equation*}
  \liminf_{N\to\infty}  N(N+\tfrac12) b_N(t)b_{N+1}(t) >0.
\end{equation*}
Unlike the case of equation~(\ref{eq:B}), if one assumes that
$\lim_{N\to\infty} b_{N+1} /b_{N}$ exists and is in $(0,\infty)$, then
the solution of equation (\ref{eq:C}) will be dissipative if
$b_n\eqsim 1/n$ as $n\to\infty$.

A solution satisfying $b_n(t) \eqsim 1/n$ (if one exists) has finite
energy: $\sum_{n=1}^\infty b^2_n(t) < \infty$. Thus, this model
differs from the first example in that the system can dissipate energy
even when the total energy is finite.

While we do not prove a general existence result as broad as for
\eqref{eq:B}, we do show (in Theorem \ref{th:c}) that the dynamics for
\eqref{eq:C} are well defined if
\begin{equation}
  \label{eq:CgoodIC}
  \sum_{n=1}^\infty (-1)^n b_n(0) < \infty.
\end{equation}
This is sufficient for our needs: In particular, it covers the case
when $b_n\eqsim 1/n$ as $n\to\infty$.

The differences between the first and second models are greater than
simply the scaling. When started with initial conditions having finite
energy the first model conserves energy. In fact the regularity at
time~$t$ is the same as the regularity of the initial condition. In
contrast, if we start \eqref{eq:C} with initial data satisfying
\eqref{eq:CgoodIC} (and hence with finite energy), the energy decays
with time. Furthermore, for almost every $t >0$ one has that
$b_n(t)\eqsim 1/n$ as $n \rightarrow \infty$ and there exists $T$,
depending on the initial data, so that if $t> s > T$ then
\begin{equation*}
  \sum_{n=1}^\infty b_n^2(t) <   \sum_{n=1}^\infty b_n^2(s) 
  <\sum_{n=1}^\infty b^2_n(T) < \infty
\end{equation*}
and $ \sum b_n^2(t) \rightarrow 0$ and $t \rightarrow \infty$.
Turning to the forced setting, consider 
\begin{equation*}
  \dot b_n = (n-1)(n-\tfrac12) b_{n-1} - n (n+\tfrac12) b_{n+1} 
  + f(t) \one_{n=1}.
\end{equation*}
When $f(t)=\dot W(t)$ then 
\begin{equation*}
  \EE b^{**}_{n} b^{**}_{n+m} \eqsim \frac1{n(n+m)}
  \quad\text{as}\quad n \rightarrow \infty.
\end{equation*}
If $f(t)=1$, we have
\begin{equation*}
  b^*_n \eqsim  \frac{1}{n} \quad\text{as}\quad n \rightarrow \infty.
\end{equation*}

\subsection{Inviscid limits of the first model}

In practice, one is often interested in understanding the limit of
equations when the explicit sources of dissipation are removed. To
explore this question we investigate equation \eqref{eq:B} with
the addition of an overtly dissipative term and then study the limit
as the dissipation is removed.

To understand our motivation, recall that we have seen that if the
first model is started with finite energy initial data then the formal
calculation presented in \eqref{eq:nodiss} is valid for all finite
times as the energy must be infinity for \eqref{eq:nodiss} to fail.
Yet, as time tends to infinity, the forced system converges to a steady
state with infinite energy for which the calculation presented in
\eqref{eq:nodiss} fails.  In contrast, in the second model the
analogous calculation, given in \eqref{eq:bNoDiss}, fails at almost
every positive time since $b(t) \eqsim 1/n$ as $n \rightarrow \infty$
for almost every $t>0$.

Since the coupling term produces dissipation at finite times in the
second model, it is most interesting to study the effect of extra,
explicit dissipation in the first model. To this end, we consider the
stochastically forced version of the first model with extra, explicit
dissipation sufficient to keep expected energy of the system finite
for all times. In particular, the calculation in \eqref{eq:nodiss} is
valid in the equilibrium state. We are interested in the structure of
this invariant state and how it converges to the steady state without
the explicit dissipation (as the dissipation is removed).

We will consider two cases, one for which the dissipative term is
lower order than the coupling term and one for which it is higher
order. Specifically, for $p \in \{0,1\}$ and any $\nu>0$, consider the
equation
\begin{equation*}
 \dot \alpha_{n,\nu} = -2\nu (n-1)^p \alpha_{n,\nu} 
 + \big[ (n-1)\alpha_{n-1,\nu}
 - n \alpha_{n+1,\nu}\big] 
 +\one_{n=1} \dot W(t). 
\end{equation*}

The case $p=0$ corresponds to the lower order damping and is analogous
to what is called Eckman damping in the context of fluid mechanics.
When $p=1$ the perturbation is higher order than the coupling term and
behaves as a viscous term in the language of fluid mechanics.  As in
the previous examples, we force the first coordinate with white noise.
Assuming that $\sum |\alpha_{n,\nu}(0)|^2 < \infty$, it is straight
forward to see that $\sum |\alpha_{n,\nu}(t)|^2$ stays finite and
uniformly bounded in time for any $\nu > 0$. Hence, the system remains
conservative for all time. Furthermore, as $t \rightarrow \infty$ the
system converges to a random variable $\alpha^{**}_{n,\nu}$ whose
distribution is the unique stationary measure for the system. Direct
calculation in the spirit of \eqref{eq:nodiss} shows that

\begin{equation*}
  \EE \sum_{n=1}^\infty (n-1)^p|\alpha_{n,\nu}^{**}|^2=\frac1{\nu}\;.
\end{equation*}
Thus, there is no anomalous dissipation in the system: All of the
dissipation which balances the forcing comes
from the term $ -2\nu(n-1)^p \alpha_n(t)$. In section
\ref{sec:invLim}, we will see that for $p \in \{0,1\}$,

\begin{equation*}
 \E [ \alpha_{n,\nu}^{**} -  a_{n}^{**}]^2 \to0 \quad\text{as}\quad \nu\to0.
\end{equation*}

Both of these steady states are Gaussian with mean zero. The way in
which the variance of $ \alpha_{n,\nu}^{**}$ converges (as
$\nu\rightarrow 0$) to that of the $\nu=0$ steady state $a_{n}^{**}$
is different for the two values of $p$ we consider. When $p=0$ one has

\begin{equation}
 \label{eq:Alphap0}
 \EE[\alpha_{n,\nu}^{**}]^2=  2^{1-2\nu}
   \Gamma(1+2\nu) \frac{\Gamma(2n+1 - 2\nu)}{\Gamma(2n +2)} 
   \eqsim\frac{1}{n^{1+2\nu}}\;.
\end{equation}
This shows that variances still decay like a power of
$n$. Notice that for all $\nu>0$ the total energy is finite.  When
$p=1$ we do not obtain an exact formula but rather that
\begin{equation}\label{eq:Alphap1}
   \frac{ \kappa^2}{(\kappa +\nu)^{2n+2}}\frac{1}{2n+1}  \leq\EE
   \big[\alpha_{n,\nu}^{**} \big]^2\leq \frac{1}{\kappa^{2n}}\frac{1}{2n+1} 
\end{equation}
where $\kappa^2=1+\nu^2$. Since $\kappa>1$ when $\nu>0$, the
$\alpha_{n,\nu}^{**}$ behave as the limiting $a_n^{**}$ for small $n$
but decay exponentially for large $n$.

\subsection{Organization}
\label{org}

The remainder of the paper is concerned with proving the statements
made in this section.  Section \ref{sec:anomalousdiss} contains a
precise discussion of the concept of anomalous dissipation. In Section
\ref{sec:examplesFirst}, we return to the first of the two models
introduced in Section \ref{sec:Mainstory} and illustrate the range of
possible dynamics by considering specific initial conditions for which
the system can be explicitly solved. In Sections \ref{sec:sol} and
\ref{sec:prop}, we prove the existence of solutions and describe their
properties for a wide range of initial data. In Section
\ref{sec:proofs}, we give the proofs of all of the preceding results.
Section \ref{sec:BForced} discusses the forced setting for the first
example and Section \ref{sec:invLim} discusses its inviscid limit.
Finally, in Section \ref{sec:general} we discuss the second model
introduced in Section \ref{sec:Mainstory}: we first describe the
qualitative behavior of solutions; then, we prove existence and
uniqueness results both in the forced and unforced situations.

We note that some of the results about anomalous dissipation and the
scaling of the solutions can also be obtained by formally taking the
continuous limit in $n$ of~(\ref{eq:B}) and~(\ref{eq:C}). In this
limit, these equation formally reduce to hyperbolic conservation laws
which were analyzed in~\cite{woodshole}.

\section{Preliminary: Definition of anomalous dissipation}
\label{sec:anomalousdiss}

The concepts of energy conservation, dissipation, and explosion are
straightforward when the total energy of the system is finite.  A
system is \textit{conservative} if the energy does not change with
time. A system is \textit{dissipative} (or \textit{explosive}) if the
total energy decreases (or increases) with time.

However, as the example in the previous section showed, it is possible
to have solutions which one might call dissipative even though the total
energy is infinite.  We give a definition of the above terms which
can be applied to situations where the total energy is infinite.

Given a time-dependent sequence, $\{a_n(t)\}_{n\in\NN}$, define the
energy in the block $M$ to $N$, $M < N$, by
\begin{equation*}
  \Eg_{M,N}(t)=\sum_{n=M}^N |a_n(t)|^2.
\end{equation*}
A given block $\Eg_{M,N}$ is \textit{dissipative} at time $t$ if $\dot
\Eg_{M,N}(t) < 0$. Similarly, we will say it is \textit{explosive} if
$\dot \Eg_{M,N}(t) > 0$. If $\dot \Eg_{M,N}(t) =0$ then we say the
block is \textit{conservative} at time $t$. If $\dot \Eg_{M,N}(t) =0$
for all $M$, $N$, and $t$, then the system is at a fixed point. (Note
that this is consistent with Example \ref{ex:fixpt} in the next
section.)

We will say that the system is \textit{locally dissipative}
(\textit{locally explosive}, or \textit{locally conservative}) at time
$t$ if every finite block is \textit{dissipative} (\textit{explosive}
or \textit{conservative}) at time $t$.

In contrast, we will say that a system with $\Eg_{0,\infty}=\infty$ is
\textit{dissipative} at time $t$ if the limit
\begin{equation*}
  \limsup_{N \rightarrow \infty} \dt \Eg_{0,N}(t)<0.
\end{equation*}
We say it is \textit{explosive} at time $t$ if
 \begin{equation*}
  \liminf_{N \rightarrow \infty} \dt \Eg_{0,N}(t)>0.
\end{equation*}
When the limit exists we will refer to its absolute value as the
\textit{rate of energy dissipation} or the \textit{rate of energy
  explosion} depending on the inequality which is satisfied.  We say
that the system with $\Eg_{0,\infty}=\infty$ is \textit{conservative}
at time $t$ if
\begin{equation*}
  \lim_{N \rightarrow \infty} \dt \Eg_{0,N}(t)=0.
\end{equation*}
If we do not state time explicitly for any of these property, we mean
that the properties holds for all finite times.

As the examples of the next section show, it is possible for
$\lim_N \Eg_{0,N}= \infty$ while $\lim_N \dot \Eg_{0,N}= c < 0$. 
\begin{remark}
It is important to notice that the above categorizations are not
exhaustive. It is possible for a system not to fit into any of the
categories. This is only an issue when the energy is infinite as we
use the definitions at the start when the energy is finite. 
\end{remark}

\section{The rich   behavior of the first model}
\label{sec:examplesFirst}

The system given by \eqref{eq:B} possesses a number of interesting
properties beyond those listed in the introductory section. In this
section we explore the behavior through a number of examples. A
relatively complete theory of the equation will be given in the two
sections which follow. We begin with an existence result which covers
all of the examples presented.

For an infinite vector $a=(a_1,a_2,\cdots)$ with $a_i \in \RR$, define
$\rho(a)$ by
\begin{equation}
  \label{eq:xi}
  \frac{1}{\rho(a)}=\limsup_{n \rightarrow \infty} |a_n|^{1/n}\;.
\end{equation}
($\rho(a)$ is simply the radius of convergence of the power series
$\sum_n a_n z^n$).  The following theorem gives an existence and uniqueness
result for \eqref{eq:B} sufficient for present needs. In particular, it
allows initial data with infinite energy ($\sum |a_n(0)|^2 = \infty$).
A more complete
description will be given in Theorem \ref{thm:Bexistance} of Section
\ref{sec:sol}, where we describe the methodology for solving
\eqref{eq:B}.

\begin{theorem}\label{thm:simpleExistence}
If  $a(0)=(a_1(0),a_2(0),\cdots)$ is an infinite vector of initial
conditions such that $\rho(a(0)) >0$, then there exists a unique
solution $a(t)$ to \eqref{eq:B} with initial conditions $a(0)$ which
exists at least up to the time $t_*=\arctanh(\rho(a)\wedge 1)$.
\end{theorem}
This existence result, whose proof is given in Section\ref{sec:sol},
covers a wide class of initial data. The dynamical behavior of
\eqref{eq:B} is quite rich. We list a number of exact solutions
which display the range of possible behaviors. Explanations of how
these results are obtained will be given in section
\ref{sec:solExamples}. A general discussion of the qualitative
properties of solutions of \eqref{eq:B} will be given in Section
\ref{sec:prop}.

\begin{example}\label{ex:conserve}
  \textbf{An energy conserving pulse heading out to infinity:} {\rm
    Fixing $a_1(0)=1$ and $a_n(0)=0$ for all $n=2,3,\ldots$ results in
    the dynamics
    \begin{equation*}
      a_n(t)=\frac{\tanh^{n-1}(t)}{\cosh(t)}.
    \end{equation*}
    Even though the solution decays to zero pointwise in $n$ as
    $t\to\infty$ it conserves energy: $\sum_{n=1}^\infty|a_n(t)|^2 =
    \sum_{n=1}^\infty|a_n(0)|^2=1$ for all $t\ge0$. The fact that it conserves
    energy is consistent with the observations in equation \eqref{eq:nodiss}
    and \eqref{eq:yesConserve} because
    \begin{equation*}
      \frac12\dt \sum_{n \le N} |a_n(t)|^2=-N a_N(t) a_{N+1}(t) = 
      -\frac{N\tanh^{N-1}(t)\tanh^N(t)}{\cosh^2(t)} \to 0 
    \end{equation*}
    as $N\to\infty$. The dynamics of this solution can be understood as
pulse moving out to larger and larger $N$ with time while
simultaneously spreading out. A simple calculation shows that $a_N(t)$
reaches its maximum at a time asymptotic to $\frac12\log(1+4N)$ as $N
\rightarrow \infty$. Hence, as $t \rightarrow \infty$ the $n$ for which
$a_n$ which is cresting at time $t$ scales as $\frac14\exp(2t)$. }
\end{example}

\begin{example}\label{ex:finDisRate}
  \textbf{Dissipative solution with finite dissipation rate:}
  {\rm If 
    \begin{displaymath}
      a_{n+1}(0)=\frac{(2n)!}{2^{2n}(n!)^2}\sim \frac{1}{\sqrt{\pi n}}\quad 
      \text{for $n=0,1,\ldots$}
    \end{displaymath}
    then
    \begin{displaymath}
      a_{n+1}(t)= \frac{e^{-t/2}}{\sqrt{\cosh(t)}}
      \sum_{m=0}^n \frac{(2(n-m))! (2m)!}{2^{2n} ((n-m)! m!)^2} \tanh^{n-m}(t)
      \quad \text{for $n=0,1,\ldots$}
    \end{displaymath}
    For each  fixed $n$, we have
    \begin{displaymath}
      a_{n}(t) \sim \sqrt{2}e^{-t} \qquad \text{as $t\to\infty$},
    \end{displaymath}
    so that the solution decays to zero pointwise in $n$ as
    $t\to\infty$. On the other hand, since for any fixed time $t$
    \begin{displaymath}
      a_{n}(t) \sim  \frac{e^{-t/2}}{\sqrt{\cosh t}} \frac{1}{\sqrt{\pi n}}
      \qquad \text{as $n\to\infty$},
    \end{displaymath}
    it follows that $\sum_{n=1}^\infty |a_n(t)|^2=+\infty$ for all time $t\ge
    0$ and 
    \begin{displaymath}
      \frac12\dt \sum_{n \le N} |a_n(t)|^2=-N a_N(t) a_{N+1}(t)  
      \to -\frac{e^{-t}}{\pi\cosh t} < 0
    \end{displaymath}
    as $N\to\infty$. For this solution, the calculation
    in~(\ref{eq:nodiss}) does not hold and the above inequality can be
    interpreted as a form of anomalous dissipation with finite
    dissipation rate.}
\end{example}

\begin{example}\label{ex:infDisRate}
  \textbf{Dissipative solution with infinite dissipation rate:}
  {\rm If $a_n(0)=1$ for all $n=1,2,\ldots$.  then
    \begin{displaymath}
      a_n(t)=e^{-t} \quad \text{for all $n=1,2,\ldots$}
    \end{displaymath}
    This solution decays to zero pointwise in $n$ as $t\to\infty$ and 
    $\sum_{n=1}^\infty|a_n(t)|^2 =+\infty$ for all
    $t\ge0$. Notice that
    \begin{displaymath}
      \frac12\dt \sum_{n \le N} |a_n(t)|^2=-N a_N(t) a_{N+1}(t) = 
      -Ne^{-2t} \to -\infty
    \end{displaymath}
    as $N\to\infty$. The formal calculation in \eqref{eq:nodiss} does
    not hold for this solution and, in terms of the definitions of
    Section~\ref{sec:anomalousdiss}, we view this as a form of
    anomalous dissipation with infinite dissipation rate.}
\end{example}

\begin{example}\label{ex:fixpt}\textbf{A fixed point:}
  {\rm If $a_{2n}(0)=0$ for $n=1,2,\dots$ and
    \begin{displaymath}
      a_{2n+1}(0)=\frac{(2n)!}{2^{2n}(n!)^2}\sim\frac{1}{\sqrt{\pi n}}\quad 
      \text{for $n=0,1,\ldots$},
    \end{displaymath}
    then
    \begin{displaymath}
      a_n(t)=a_n(0) \quad \text{for all $n=1,2,\ldots$}
    \end{displaymath}
    This solution is a fixed point of~(\ref{eq:B}). Notice that
    \begin{displaymath}
      \frac12\dt \sum_{n=1}^N |a_n(t)|^2= - N a_N(t) a_{N+1}(t) = 0
    \end{displaymath}
    for all $N\in\NN$, but $\sum_{n=1}^\infty|a_n(t)|^2 =+\infty$ since
    $a_{2n+1}\sim1/\sqrt{\pi n}$ as $n\to\infty$.
  }
\end{example}

\begin{example}\label{infExpTime}\textbf{Explosive solution with infinite explosion time:}
  {\rm If $a_{n}(0)= (-1)^{n+1}$ for $n=1,2,\dots$.  then
    \begin{displaymath}
      a_n(t)= (-1)^{n+1} e^t \quad \text{for all $n=1,2,\ldots$}
    \end{displaymath}
    In this case, $\sum_{n=1}^\infty|a_n(t)|^2 =+\infty$ and 
    \begin{displaymath}
      \frac12\dt \sum_{n=1}^N |a_n(t)|^2= - N a_N(t) a_{N+1}(t) = N e^{2t}  
      \to +\infty
    \end{displaymath}
    as $N\to\infty$.  Thus, (\ref{eq:nodiss}) does not hold for this
    solution and we see that the above limit is
    consistent with an infinite explosion rate for the solution. This
    is consistent with Theorem \ref{thm:simpleExistence} as
    $\rho(a)=1$ and $t_*=\arctanh(1)=\infty$ }.
\end{example}

\begin{example}\label{ex:finExpTime}\textbf{Explosive solution with finite explosion time:}
  {\rm If $a_{n}(0)= (-1)^{n+1}\alpha^n $ with $\alpha>1$ for
    $n=1,2,\dots$,
    then
    \begin{equation*}
      a_n(t)= \frac{(-1)^{n+1}\alpha^n }{\cosh(t) - \alpha \sinh(t)} 
      \quad \text{for $t < t_* = \arctanh (1/\alpha)$ and all $n=1,2,\ldots$}
    \end{equation*}
    This solution blows up at $t=t_*$.  In this case, $\sum_{n=1}^\infty|a_n(t)|^2
    =+\infty$ for all $t<t_*$ and
    \begin{equation*}
      \frac12\dt \sum_{n \le N} |a_n(t)|^2= - N a_N(t) a_{N+1}(t) =
      \frac{N\alpha^{2N+1}}{\big(\cosh( t) - \alpha \sinh(t)\big)^2}
      \to +\infty
    \end{equation*}
    as $N\to\infty$ for all $t<t_*$.  Notice that this example is
    consistent with Theorem \ref{thm:simpleExistence} as
    $\rho(a)=1/\alpha <1$.}
\end{example}

The above examples demonstrate the rich range of behavior of the
model. In particular, some solutions grow coordinate-wise in time
while others decay. The following result gives a criteria for the later.

\begin{theorem}\label{thm:consEnergy}
  Let $(a_1(0),a_2(0),\cdots)$ be the infinite vector of initial
  conditions.  If the limit 
  \begin{equation}\label{eq:abelLimit}
    \lim_{r \rightarrow -1^+}\sum_{n=1}^\infty a_n(0)r^n
  \end{equation}
  exists and is finite, then for all $n\in \NN$, $|a_n(t)| \rightarrow
  0$ as $t \rightarrow \infty$.
\end{theorem}

Looking back at the examples, this result correctly separates those
which decay to zero pointwise in $n$ and those which do not.

Denote by $\ell^p$  the $p$-summable sequences: For
$p > 0$,
\begin{equation}\label{eq:lp}
  \ell^p := \big\{ (a_1,a_2,\ldots): \sum_{n=1}^\infty |a_n|^p < \infty\big\}.
\end{equation}
If $a=(a_1,a_2,\ldots) \in \ell^1$ then \eqref{eq:abelLimit}
exists and is finite. For future reference, we recall the norm
$\|\cdot\|_{\ell^p}$ defined by $\|a\|^p_{\ell^p} \eqdef
\sum_{n=1}^\infty |a_n|^p$.

A complimentary question is to understand for which initial data the
system conserves energy.
\begin{theorem}\label{thm:conserveEnergy}
  If $a(0)=(a_1(0),a_2(0),\cdots) \in \ell^2$ , then 
  \begin{align*}
    \|a(0)\|^2_{\ell^2}= \|a(t)\|^2_{\ell^2}
  \end{align*}
for all time $t \geq 0$.
\end{theorem}

Example \ref{ex:conserve}, \ref{ex:finDisRate} and \ref{ex:infDisRate}
all have nice limits at $z=-1$ in the sense of \eqref{eq:abelLimit} and they
decay to zero as $t \rightarrow \infty$ as dictated by Theorem
\ref{thm:consEnergy}.  It is particularly interesting to compare
Example \ref{ex:infDisRate} and \ref{infExpTime}. Theorem
\ref{thm:consEnergy} correctly says that the first decays to zero as time
increases while declining to comment on the second.

Theorem \ref{thm:conserveEnergy} correctly states that Example
\ref{ex:conserve} conserves energy. However, Theorem
\ref{thm:conserveEnergy} is not completely satisfactory in that it
only applies to solutions which have finite total energy. A number
of our example have infinite energy. We now turn to understanding in
detail the dynamics of \eqref{eq:B}.

\section{Solution to the initial value problem}
\label{sec:sol}

In this section we show that the initial value problem associated to
equation~\eqref{eq:B} is well-posed and admits solutions for a wide
class of initial data. Theorem~\ref{thm:simpleExistence} above is an
immediate consequence of Theorem~\ref{thm:Bexistance} below. After
giving a general existence and uniqueness theorem, we present specific
initial conditions (covered by the existence theorem) for which
equation~\eqref{eq:B} admits solutions which conserve energy,
dissipate energy, and blow up in finite time.

We begin by formally calculating a representation of the solution
given by a generating function. We will verify that the representation
is valid in the next section.  Given initial conditions $\{ a_n(0) :
a_n(0) \in \RR, n \in\NN \}$, we assume that a solution $a_n(t)$
exists and define the generating function
\begin{align}\label{eq:Gt}
G(z,t) = \sum_{n =0}^\infty a_{n+1}(t)z^n\;.
\end{align}
Proceeding formally, it is straight forward to verify that $G(z,t)$
would satisfy the following partial differential equation:
\begin{align}
  \label{eq:GeqForB}
  \frac{\partial G}{\partial t}= (z^2 -1) \frac{\partial G}{\partial
    z} + z G
\end{align}
with initial condition
\begin{align}\label{eq:G0}
G_0(z) \eqdef  \sum_{n =0}^\infty a_{n+1}(0)z^n\;.
\end{align}
The first term on the right hand side of  \eqref{eq:B} comes from 
$z \frac{\partial \ }{\partial z} (z G)$ and the second from
$-\frac{\partial \ }{\partial z}G$.

One obtains an ansatz for the form of the solution by solving
equation~\eqref{eq:GeqForB} by the method of characteristics. By
verifying that this ansatz solves the equation, we obtain the
following existence and uniqueness result whose proof is postponed
until section \ref{sec:proofs}.
\begin{theorem}
  \label{thm:Bexistance}
  Consider~(\ref{eq:B}) with the initial condition $a_n(0)$ such that
  $G_0(z)$ is analytic in a neighborhood of the interval
  $(\alpha,0]$. Then the solution of~(\ref{eq:B}) exists and is unique
  for all $t \in [0,t_*)$ where $t_*=\arctanh(-\alpha \wedge 1)$.

  The unique solution is given by 
\begin{align}  \label{eq:anReconstruction}
  a_n(t) = \oint_\Gamma \frac{G(z,t)}{2\pi i z^{n}} dz
\end{align}
with 
\begin{equation}
  \label{eq:Gsol}
    G(z,t) \eqdef \frac{\psi_t(z)}{\cosh(t)} (G_0\circ\phi_t)(z),
\end{equation}
where
\begin{equation}
  \label{eq:defPsiPhi}
  \psi_t(z)\eqdef\frac{1}{1- z \tanh(t)} \quad\text{and}\quad
  \phi_t(z)\eqdef\frac{z-\tanh(t)}{1- z \tanh(t)}. 
\end{equation}
$\Gamma$ is any simple closed contour around the origin within the
region of analyticity of $G(z,t)$ (which is non-empty). 
\end{theorem}
It is worth noting that $G(z,t)$ solves the PDE given in
\eqref{eq:GeqForB} with $G(z,0)=G_0(z)$ as initial condition.

Notice that finite time existence of solutions only requires that the
initial data $a_n(0)$ have at most exponential growth in $n$, i.e.
there exists $C>0$ and $\gamma>1$ such that for all $n\in\NN$,
$|a_n(0)|\le C \gamma^n$. If the $a_n(0)$ decay exponentially then the
solution exists for all times (i.e. $t_*=\infty$).

In addition, (\ref{eq:anReconstruction}) implies that $a_{n+1}(t)$ is
the $n$th term of the Taylor series expansion of $G(z,t)$ about $z=0$.
It is also straight forward to see that (\ref{eq:anReconstruction})
defines a semigroup: for any suitable $f(z)$, let
\begin{equation}
  \label{eq:semigroup}
  (S_tf)(z) := \frac{1}{\cosh(t)} \psi_t(z) f\big(\phi_t(z) \big)\;.
\end{equation}
Then (\ref{eq:Gsol}) can be expressed as $G(\cdot,t)= S_tG_0$ and it
is easy to check that for any $t,s>0$,
\begin{equation*}
  S_t\circ S_s f =  S_s\circ S_t f = S_{t+s} f.
\end{equation*}
Since we are particularly interested in knowing the total energy of
the solution, it is useful to notice that if $G(x,t)$ is as in
\eqref{eq:Gt} then
\begin{align*}
  \|a(t)\|^2_{\ell^2}\eqdef \sum_{n=1}^\infty |a_n(t)|^2 =
 \frac1{2\pi} \int_0^{2\pi} |G(e^{i \theta},t)|^2 d\theta \eqdef \|G(e^{i
    \theta},t)\|^2_{L^2(S^1,d\theta)}\;.
\end{align*}

We shall give more properties of the solutions of equation~\eqref{eq:B} in
section~\ref{sec:prop} after a brief discussion of the examples given in
the previous section.

\subsection{Analysis of examples}
\label{sec:solExamples}
We use Theorem \ref{thm:Bexistance} to calculate exact the
solutions given in Section \ref{sec:examplesFirst}.

The initial data in Example \ref{ex:conserve}, translates in to
$G_0(z) = 1$, so that
\begin{equation*}
G(z,t)=\frac{1}{\cosh(t)-z \sinh(t)}\;.  
\end{equation*}
Calculating the Taylor series gives the quoted $a_n(t)$.
In example \ref{ex:finDisRate} one obtains $G_0(z) = 1/\sqrt{1-z}$. Hence,
\begin{equation*}
  G(z,t)=\frac{e^{-t/2}}{\sqrt{\cosh(t)}} \frac1
    {\sqrt{(1-z)(1-z\tanh(t))}}\;,
\end{equation*}
whose Taylor expansion produces the quoted $a_n(t)$.
Example \ref{ex:infDisRate} yields $G_0(z) = 1/(1-z)$,
$G(z,t)=e^{-t}/(1-z)$ and the desired $a_n$.  Example \ref{ex:fixpt}
gives $G_0(z) = 1/\sqrt{1-z^2}$ and $G(z,t)=1/\sqrt{1-z^2} \equiv
G_0(z)$.  Example \ref{infExpTime} gives $G_0(z) = 1/(1+ z)$ and
$G(z,t)=\frac{ e^t }{1+z}$.  Example \ref{ex:finExpTime} gives $G_0(z)
= \alpha /(1+ \alpha z)$ and
\begin{equation*}
  G(z,t)=\frac1{\cosh(t) - \alpha \sinh(t)}\frac{\alpha   }{1+\alpha z} \qquad
    \text{ for $t < t_* = \arctanh (1/\alpha)$};
\end{equation*}
it blows up at $t=t_*$ as stated.

\section{Properties of the solutions}
\label{sec:prop}

We begin by presenting two results which are more quantitative
versions of the results in Theorem \ref{thm:consEnergy} and Theorem
\ref{thm:conserveEnergy}.  Together, they highlight the fact that it
is possible to have a given coordinate converge to zero while no
global energy dissipation is present in the system. This implies that
there is a flux of energy out to higher and higher modes.  It is also
interesting that both of the next two results apply in some situations
where the total energy is infinite.

\begin{theorem}
\label{thm:gozero}
Suppose that $G_0(z)$ is analytic in a neighborhood of  $(-1,0]$ such that 
\begin{equation}
  \label{eq:G0Limit}
  G_0^+(-1)=\lim_{\substack{x\rightarrow -1^+\\ x\in \RR}} G_0(x)
\end{equation}
exists and is finite. Then, 
  \begin{equation*}
    a_n(t) \sim 2 e^{-t} G_0^+(-1) \quad \text{as $t\to\infty$}.
  \end{equation*}
In particular, in a neighborhood of the origin $G(z,t) \sim 2
e^{-t}G_0^+(-1)/(1-z)$ as $t \rightarrow \infty$.
\end{theorem}
\begin{remark}
  If $a_n=n^{-\alpha}$ for $\alpha < 1$, then $a_n(t) \sim 2 e^{-t}
  \Gamma(1-\alpha)$ as $t\to\infty$. 
\end{remark}
The next result contains Theorem \ref{thm:conserveEnergy} as well
as giving control of  higher Sobolev-like norms. We recall 
the Sobolev-like sequence spaces for $s \in \RR$:
\begin{equation*}
  h_s\eqdef\left\{ a=(a_1,a_2,\ldots) : \|a\|_{h_s} < \infty\right\}
\end{equation*}
where the norm $\|\cdot\|_{h_s}$ is defined by $\|a\|_{h_s}^2 \eqdef
\sum_{n=1}^\infty n^{2s}|a_n|^2$.
\begin{theorem}\label{thm:sobbound}
  If 
  \begin{equation*}
    \|a(0)\|^2_{\ell^2}\eqdef\sum_{n=1}^\infty |a_n(0)|^2 < \infty
  \end{equation*}
then 
  \begin{equation*}
    \|a(0)\|^2_{\ell^2}=\|a(t)\|^2_{\ell^2}
  \end{equation*}
for all $t\geq 0$. Similarly,  for any $s \in \NN$, if 
  \begin{align*}
    \|a(0)\|_{h_s}^2 \eqdef\sum_{n=1}^\infty n^{2 s} |a_n(0)|^2 < \infty ,
  \end{align*}
  then for any $T<\infty$ there exists a constant, $C(T)$, such that
  \begin{align*}
    \sup_{t\in[0,T]}\|a(t)\|_{h_s}^2 \leq C(T).
  \end{align*}
\end{theorem}

\subsection{Finer properties of solutions}
We begin by giving conditions guaranteeing that the solution decays
exponentially in time. We then turn to the case for which $G_0$ has
singularities on the boundary of the unit circle. We close the section
with a result which compares  the dynamics obtained by placing a
single unit of mass at different locations.

Let $D(r)$ denote the open disk of radius $r$ about the origin. We
begin with a simple criteria which guarantees that the $a_n$ decay
exponentially in $n$.
\begin{theorem}
\label{thm:conservesAnalytic}
Suppose $G_0(z)$ is analytic in the disk $D(1+\eta)$ for some $\eta >
0$. Then, for each $T \geq 0$, there exist constants $\gamma, C>0$
which depend only on $\eta$ and $T$ such that
\begin{equation*}
  \sup_{t\in [0,T]}|a_n(t)| \leq C \gamma^n \text{ for all $n \in \NN$}.
\end{equation*}
In particular, the system conserves energy at all finite times.
\end{theorem}

In order to investigate the behavior of solutions when $G_0(z)$ is not
analytic in $D(1+\eta)$ for some $\eta>0$, we introduce the following
region:
\begin{equation}\label{eq:detlaRegion}
  \Delta(\zeta,\eta,\theta)\eqdef\big\{ z : |z| 
  \leq |\zeta|+\eta, |\arg(z-\zeta)-\arg(\zeta)| \geq \theta\big\}\;.
\end{equation} 
We begin with a careful analysis of the case when there is a single
singularity on the unit circle. We contrast the cases for which the
singularity is located at $\pm 1$ in Theorem \ref{th:2} and Theorem
\ref{th:3}, respectively. The remaining cases are covered by Theorem
\ref{th:remainderOfDisk}.

\begin{theorem}\label{th:2}
Assume that either
\begin{equation*}
  a_n(0)\sim  Cn^{\alpha-1} \quad \text{as $n\to\infty$}
\end{equation*}
for some $C \neq 0$ or, more generally, that $G_0(z)$ satisfies both of
the following conditions:
\begin{enumerate}
\item There exist an  $A\neq 0$\ and\ $\alpha>0$ so that
  \begin{equation}
    \label{eq:G0blow2}
    G_0(z) \sim A (1-z)^{-\alpha} \quad 
    \text{  as\  $z\to 1$.}
  \end{equation}
\item $G_0(z)$ is analytic on $\Delta(1,\eta,\theta)\backslash\{1\}$
  for some $\eta>0$ and $0< \theta< \pi/2$.
\end{enumerate}
Then, for all time $t\ge 0$, 
\begin{equation}
  \label{eq:decayalpha}
  a_n(t)\sim
  \frac{A}{\Gamma(\alpha)}e^{(1-2\alpha)t}
  n^{\alpha-1}
  \quad \text{as $n\to\infty$,}
\end{equation}
and the solution decays to zero as $t\to\infty$ pointwise
in $n$; more precisely,
\begin{equation*}
  a_n(t) \sim  A 2^{1-\alpha}e^{-t} \quad \text{as $t\to\infty$}\;.
\end{equation*}
\end{theorem}
Notice that in the setting of Theorem \ref{th:2} the energy of the
system is infinite for all $\alpha>0$: $\sum_{n=1}^\infty a_n^2(t) =
+\infty$. However, as a direct consequence of (\ref{eq:decayalpha}) in
Theorem~\ref{th:2} and our definition of dissipation in
section~\ref{sec:anomalousdiss} the following is true:
\begin{corollary}
  \label{cor:1}
  In the setting of Theorem \ref{th:2}: if $ 0< \alpha <1/2$ the system
  is conservative; if $\alpha = 1/2$ the system displays a finite dissipation
  rate; and, if $\alpha > 1/2$ the system displays an infinite dissipation
  rate.
\end{corollary}

We now consider a singularity at $z=-1$. The remaining points on the
unit circle behave much like the $z=1$ in that the solution decays to
zero in time. They are discussed in Theorem \ref{th:remainderOfDisk}
later in the section. The next result shows that if there is a
singularity at $z=-1$ then the system can explode in time.

\begin{theorem}
  \label{th:3}
Assume that either
\begin{equation*}
  a_n(0)\sim  C (-1)^n n^{\alpha-1} \quad \text{as $n\to\infty$}
\end{equation*}
for some $C \neq 0$ or, more generally, that $G_0(z)$ satisfies both of
the following conditions:
\begin{enumerate}
\item  There exist an  $A\neq 0$\ and\ $\alpha>0$ so that
\begin{equation}
  \label{eq:G0blow3}
  G_0(z) \sim A (z+1)^{-\alpha} \quad 
  \text{ as\  $z\to -1$.}
\end{equation}
\item  $G_0(z)$ is analytic on $\Delta(-1,\eta,\theta)\backslash\{-1\}$
  for some $\eta>0$ and $0< \theta< \pi/2$.
\end{enumerate}
Then, for all time $t\ge 0$, 
\begin{equation*}
  a_n(t)\sim \frac{A}{\Gamma(\alpha)}e^{(2\alpha-1)t} (-1)^n n^{\alpha-1} 
  \quad \text{as $n\to\infty$}
\end{equation*}
while for $n$ fixed
\begin{equation*}
  a_n(t) \sim 2^{1-\alpha} e^{(2\alpha-1)t} A C_{n-1}^\alpha \quad 
  \text{as $t\to\infty$},
\end{equation*}
where $C_n^\alpha$ is the $n^{th}$ coefficient of the Taylor series
expansion at $z=0$ of the function
\begin{equation*}
  \frac{1}{1-z}\left[\frac{1-z}{1+z}\right]^\alpha\;.
\end{equation*}
\end{theorem}

Notice that the energy of the system is again infinite for all
$\alpha>0$: $\sum_{n=1}^\infty |a_n(t)|^2 = +\infty$. However, we
have:
\begin{corollary}
  \label{cor:2} In the setting of Theorem \ref{th:3}: If $0<
  \alpha <1/2$, the system is conservative; if $\alpha = 1/2$, it
  displays a finite explosion rate; and, if $\alpha > 1/2$, it displays
  an infinite explosion rate.
\end{corollary}

\begin{remark} One must be careful when interpreting the results in
  Corollary~\ref{cor:2} since the system has infinite energy. For
  example, when $\alpha = 1/2$ then at each moment of time the system
  displays a finite explosion rate since energy is pumped in from
  infinity into any finite collection of modes. However, the rate must
  slow, falling to zero at $t=\infty$, since as $t \rightarrow \infty$
  the system converges to the fixed point given in
  Example~\ref{ex:fixpt}.  This follows from the fact discussed in
  Section~\ref{sec:solExamples} that the fix point in
  Example~\ref{ex:fixpt} corresponds to the initial function
  $G(z)=B/\sqrt{1-z^2}$ for some $B \neq 0$.
\end{remark}

We now give a more general result covering a singularity on the unit
circle at any point other than $-1$. Theorem \ref{th:2} is a special
case of the following result when $\zeta=1$.
\begin{theorem}\label{th:remainderOfDisk} 
  Let $\zeta$ be a point on the unit circle not equal to $-1$.  Assume
  that $G_0(z)$ behaves as
\begin{equation}
  \label{eq:G0blow4}
  G_0(z) \sim A (\zeta-z)^{-\alpha} \quad 
  \text{with\ $A\neq 0$\ and\ $\alpha>0$\  as\  $z\to \zeta$},
\end{equation}
and that $G_0(z)$ is analytic on $
\Delta(\zeta,\eta,\theta)\backslash\{\zeta\}$ for some $\eta>0$ and $0<
\theta< \pi/2$. Then, for all time $t\ge 0$, 
\begin{equation*}
  a_n(t)\sim
  \frac{A}{\Gamma(\alpha)}\left[\left(\frac{1+\zeta}{2}\right)e^t + 
    \left(\frac{1-\zeta}{2}\right)e^{-t} \right]^{1-2\alpha}\zeta_t^n
  n^{\alpha-1}
  \quad \text{as $n\to\infty$}
\end{equation*}
where $\zeta_t=\phi_t^{-1}(\zeta)$. (Notice that $\phi_t^{-1}(1)=1$
and $\zeta_t \rightarrow 1$ as $t\rightarrow \infty$). The solution
decays to zero as $t\to\infty$ pointwise in $n$:
\begin{equation*}
  a_n(t) \sim  2e^{-t}\frac{A}{(1+\zeta)^\alpha} \quad \text{as $t\to\infty$}\;.
\end{equation*}
\end{theorem}

From the examples above it is natural to conjecture that any
singularity on the unit circle dominated by a polynomial-like
singularity of degree less than $1/2$ will not destroy energy
conservation. It can be shown that this intuition is correct
for a wide class of initial conditions. We already know that if the
initial conditions have finite energy, then energy is conserved.  By
the reasoning in the section defining anomalous dissipation, it is
enough to have
\begin{equation}\label{eq:halfdecay}
   \lim_{n\rightarrow \infty} n^\frac12 |a_n| = 0\;.
\end{equation}
This is possible even if the total energy is infinite. Using the
Tauberian theorems in \cite{Hilberdink01TPS} we can show that the
dynamics preserves a subset  of sequences satisfying
\eqref{eq:halfdecay}. These solutions  have infinite energy yet
conserve energy in the sense of  Section \ref{sec:anomalousdiss}. As
these results are tangential and a bit technical we do not give the details.

\subsection{The fixed point}
\label{sec:fixedpoint}

By combining the results of the previous section, one can understand a
wide range of behavior. We illustrate this by examining the
convergence to a fix point.  In Example~\ref{ex:fixpt}, we saw that
the initial data corresponding to $G_0(z)=1/\sqrt{1-z^2}$ was
invariant under the dynamics. This function is analytic in the open
unit disk and has two square root singularities on the unit circle:
one at $z=1$ and another at $z=-1$. Up to a technicality, we show that
this is the only fixed point that characterizes the initial data that
converges to it.

If the solution is to exist for all times the function $G_0(z)$ must
be analytic in a neighborhood of $(-1,0]$. From
Theorem~\ref{thm:gozero}, we see that if the limit as $z \rightarrow
-1$ of $G_0(z)$ is finite then the solution converges to zero
pointwise in $n$ as $t\rightarrow \infty$. Hence, if the system
converges to a nontrivial steady state, it must have a singularity at
$z=-1$. We assume that the singularity is power-like (i.e.
$(z+1)^{-\alpha}$). One can likely deal with other singularities,
however, we choose not to pursue this matter here.  Theorem~\ref{th:3}
implies that the singularity must be order $1/2$ if it is a power;
otherwise, the system would blow up or decay to zero. The question is:
Can one make sense of the dynamics when $\sum |a_n|^2 = \infty$?  The
two facts above imply that any initial condition which has only
polynomial singularities at $z=-1$ (if any) and which converges to a
fixed point must be of the form:
\begin{equation}\label{eq:fixptICform}
  G_0(z)=\frac{A}{(1+z)^\frac12}+ \tilde G_0(z) 
\end{equation}
where $A\neq 0$ and $\tilde G_0$ is analytic in a neighborhood of
$(-1,0]$ and $\tilde G_0(z) \rightarrow \tilde G_0^+< \infty$ as $z
\rightarrow -1$ from the right along the real axis. By
Theorem~\ref{thm:gozero} the dynamics starting from $\tilde G_0$
converge to zero as $t \rightarrow \infty$ pointwise in the sequence
space. Theorem~\ref{th:2} says that the first term converges to
$A\sqrt{2/(1-z^2)}$ as $t\rightarrow \infty$. Hence, all data of the
form \eqref{eq:fixptICform} converge to the fix point from
Example~\ref{ex:fixpt}. We have proved the following result:
\begin{theorem}\label{thm:Fixpt} Let $(a_1(0),a_2(0),\ldots)$ be
  an initial condition such the solutions exists for all time and
  converges coordinate-wise to a fixed vector $(\bar a_1,\bar
  a_2,\ldots)$ which is not the zero vector. Assuming that the $G_0(z)$
  associated to this initial condition has only a power-like singularity at
  $z=-1$ then $G_0$ is of the form given in
  \eqref{eq:fixptICform} and
\begin{equation}\label{eq:fixptform}
  \sum_{n=0}^\infty \bar a_{n+1} z^n= A\sqrt{\frac{2}{1-z^2}}\;.
\end{equation}
In particular, all fixed points of equation \eqref{eq:B} are given by
the Taylor series of~\eqref{eq:fixptform}.
\end{theorem}

Notice that the fixed points have infinite energy and have the
property that all of the $a_n$, with $n$ odd, are equal to zero. In
Section~\ref{sec:BForced}, we will see that steady states with less a
degenerate structure are obtained by forcing the system.

\begin{remark} If one starts with an initial condition which has a
  polynomial singularity on the unit circle at $\zeta \neq -1$ and of
  an order $\beta+1/2$ with $\beta > 0$, then for all finite times the
  norm $\|a(t)\|_{h_s}$ will be infinite if $s \geq -\beta$. Yet the
  dynamics still converges to the fix point given by Theorem
  \ref{thm:Fixpt}. Hence, at $t=\infty$ all $\|\cdot \|_{h_s}$ norms
  are finite for $s< 0$.
\end{remark}

\subsection{The effect of shifting the initial condition}
\label{sec:shiftfource}

In Example \ref{ex:conserve} we described how a unit of mass placed at
$a_1$ spreads out. The following theorem states that the solution
obtained by placing a unit of mass in $a_p$ behaves in the same way as
the solution obtained by placing a mass at $a_1$ except that the
picture is shifted $p$ units down the chain.

\begin{theorem}\label{thm:powers}
  Let $G_0(z)=z^p$ for some $p \in \NN$, then
  \begin{equation*}
    a_n(t)= (-1)^p \frac{\tanh^{n+p}(t)}{\cosh(t)} + \beta_n(t)
    \frac{\tanh^{n-p+\xi}(t)}{\cosh(t)}   
  \end{equation*}
where $\xi=p-p\wedge n$ and  $\beta_n(t)$ satisfies
\begin{equation*}
  |\beta_n(t)| \leq
  \begin{cases}
    \;n^p \big[1 - \tanh^{2p}(t) \big] & n \geq p\\
    \;n^p \tanh^{p-n}(t)\big[1 - \tanh^{2n}(t) \big] & n < p
  \end{cases}
\end{equation*}
\end{theorem}
\begin{proof}[Theorem \ref{thm:powers}]
  First observe that 
    \begin{equation*}
      G(z,t)=\frac{\cosh^{-1}(t)}{1-z \tanh(t)}
      \left[\frac{z-\tanh(t)}{1-z \tanh(t)} \right]^p.
    \end{equation*}
Expanding this we find
\begin{equation*}
  G(z,t)= \frac{1}{\cosh(t)}\sum_{n=0}^\infty z^n \big[ (-1)^p
  \tanh^{n+p}(t) + \beta_n\tanh^{n-p+\xi}(t) \big]   
\end{equation*}
where 
\begin{align*}
  \beta_n(t) &=\tanh^{p\wedge n}(t)\sum_{k=1}^{p\wedge n}
  \left[\tanh^{-1}(t)-\tanh(t)\right]^k \left[-\tanh(t)\right]^{p-k} \left(
    \begin{matrix}
      p \\
      k
    \end{matrix}\right)
\left(
    \begin{matrix}
      n \\
      k    \end{matrix}\right). \\
  \intertext{So} |\beta_n(t)| &\leq n^p \sum_{k=1}^{p\wedge n}
  \big[1-\tanh^2(t)\big]^k \big[\tanh(t)\big]^{p\wedge n-k}
  \big[\tanh(t)\big]^{p-k} \left(
    \begin{matrix}
      p \\
      k
    \end{matrix}\right).
\end{align*}
In both cases the quoted estimate follows by using the binomial theorem.
\end{proof}

\section{The forced system}
\label{sec:BForced}

Since \eqref{eq:B} may display anomalous dissipation, it is not
unreasonable to expect that adding a forcing term to this equation may
lead to a (statistical) steady state. We now 
show that this is indeed the case. Specifically, we study the system
\begin{equation}
  \label{eq:B2}
  \dot a_n(t) = (n-1)a_{n-1}(t) - n a_{n+1}(t) +\one_{n=m} f(t), 
\end{equation}
where $m\in \NN$, and $f(t)$ is either a constant forcing term, $f(t)
= 1$, or a white-noise process, $f(t) = \dot W(t)$ (in the second
case~(\ref{eq:B2}) has to be properly interpreted as an infinite
system of coupled It\^o stochastic differential equations).

As in the unforced setting, we represent the solution to~(\ref{eq:B2}) as
in~(\ref{eq:anReconstruction}) for some $G(z,t)$. $a_n(t)$ is 
 the $n$th coefficient in the Taylor series expansion of $G(z,t)$.
By Duhamel's principle one sees that $G(z,t)$ must satisfy the
generalization of~(\ref{eq:GeqForB}) with the effect of $f(t)
\one_{n=m}$ included:
\begin{displaymath}
  \frac{\partial G}{\partial t}= (z^2 -1) \frac{\partial G}{\partial
    z} + zG + F(z,t), 
\end{displaymath}
where $F(z,t) = z^{m-1} f(t)$ and the initial condition is
$G(z,0)=G_0(z)$.  This equation is valid for both $f(t)=1$ and $f(t) =
\dot W(t)$ and forcing on any $m\in\NN$.  Using the semigroup
representation defined in~(\ref{eq:semigroup}), the solution of the
equation above can be represented as
\begin{equation}
  \label{eq:}
  G(z,t) = (S_t G_0)(z) + \int_0^t (S_{t-s}F)(z,s) ds.
\end{equation}
We have

\begin{theorem}
  \label{th:f1}
  Consider~(\ref{eq:B2}) with $m=1$, $f(t)=1$, and initial condition
  $a_n(0)$ satisfying the assumptions of Theorem \ref{thm:gozero}.
  Then,
  \begin{equation}
    \label{eq:limt}
    \lim_{t\to\infty} a_n(t) =  a_n^* =
    \begin{cases}
      \displaystyle \sqrt{ \frac{\pi}2} & \text{if }\; n =1\\
      \displaystyle
      \frac{\sqrt{\pi}}{n-1}\cdot\frac{\Gamma(\frac{n}2)}{\Gamma(\frac{n-1}2)}
      & \text{if }\; n \geq 2.
    \end{cases}
  \end{equation}
\end{theorem}

\begin{remark}
  Notice that
  \begin{displaymath}
    a_n^* \sim \sqrt{\frac{\pi}{n}} \qquad \text{as $n\to\infty$}
  \end{displaymath}
  This implies that $a_n^*$ has infinite energy: $\sum_{n=1}^\infty
  |a_n^*|^2=+\infty$.  This is consistent with the fact that the
  steady state must be dissipative to compensate for the effect of the
  forcing term since dissipative solutions must have infinite energy.
  In this simple example we can see how the forcing is balanced
  explicitly. Mirroring the calculation in
  \eqref{eq:dissipationModelOne} for any $N$: If we start the system
  $\{a_n^*\}$ at time $t=0$ we have
  \begin{align*}
     \frac12\dt \sum_{n \le N} |a_n^*(t)|^2&= a_1^* - N
     a_N^* a_{N+1}^*=0
  \end{align*}
Hence, every block conserves energy as must happen at a fix point.
\end{remark}

\begin{theorem}
  \label{th:f2}
  Consider~(\ref{eq:B2}) with $m=1$, $f(t)=\dot W(t)$, and initial
  condition $a_n(-T)$ satisfying the assumptions of Theorem
  \ref{thm:gozero}.  Then,
  \begin{equation}
    \label{eq:limt2}
    \lim_{T\to\infty} a_n(t) =  a_n^{**}(t) \equiv 
    \int_{-\infty}^t\frac{\tanh^{n-1}(t-s)}{\cosh(t-s)} dW(s) \qquad {a.s.}
  \end{equation}
\end{theorem}

\begin{remark}
  From~(\ref{eq:limt2}), $a_n^{**}(t)$ is a Gaussian process with mean
  zero and covariance
  \begin{equation}
    \label{eq:cov}
    \begin{aligned}
      \EE a^{**}_n(t) a_{m}^{**}(t)& 
      = \int_{-\infty}^t \frac{\tanh^{n+m-2}(t-s)}{\cosh^2(t-s)} ds\\
      &= \frac1{n+m-1} \qquad\qquad n,m \in
      \NN.
    \end{aligned}
\end{equation}
  Again, this is consistent with the need for dissipation and 
  implies that the invariant measure for~(\ref{eq:B2}) with a
  white-noise forcing is supported on functions with infinite energy
  (and, in particular, (\ref{eq:cov}) is not trace-class). 

  In addition, notice that this is consistent with the fact
  that, at least in expectation, the steady state needs to dissipate
  precisely the energy pumped into the system. In fact, for any $N$,
 \begin{align*}
   \frac12\dt \EE\sum_{n \le N} |a_n^{**}(t)|^2&= \frac12 - N
   \,\EE\, a_N^{**}(t) a_{N+1}^{**}(t)= \frac12- \frac{N}{2N}=0\;.
  \end{align*} 
\end{remark}

\begin{proof}[Theorem~\ref{th:f1}]
The first term on the right hand-side accounts for the initial
condition. Theorem~\ref{thm:gozero} implies that
$(S_t G_0)\to0$ as $t\to\infty$. The second term is given explicitly by
\begin{equation}
  \label{eq:forcedsol}
  \begin{aligned}
    &\int_0^t (S_{t-s}F)(z,s) ds = \int_0^t
    \frac{\psi_{t-s}(z)}{\cosh(t-s)} \left(\phi_{t-s}(z)\right)^{m-1}
    f(s) ds
\end{aligned}
\end{equation}
Letting $f(t)=1$ and $m=1$, this expression becomes
\begin{displaymath}
  \int_0^t (S_{t-s}F)(z,s) ds= \int_0^t \frac{\psi_{t-s}(z)}{\cosh(t-s)} ds.
\end{displaymath}
It follows that
\begin{displaymath}
  \lim_{t\to\infty} \int_0^t (S_{t-s}F)(z,s)ds\\
    = \frac{2}{\sqrt{1-z^2}} \left(\arctan\Big(\frac{z-1}{\sqrt{1-z^2}}\Big)
      +\pi\right).
\end{displaymath}
$a_n^*$ is the $n$th coefficient of the Taylor series expansion at
$z=0$ of this function.
\end{proof}

\begin{proof}[Theorem~\ref{th:f2}]
  Letting $f(t)=\dot W(t)$, $m=1$, and
  considering the initial condition at  $t=-T$, we have
\begin{displaymath}
  \int_{-\infty}^t (S_{t-s}F)(z,s) ds= \int_{-\infty}^t  
  \frac{\psi_{t-s}(z)}{\cosh(t-s)} dW(s).
\end{displaymath}
$a^{**}(t)$ is the $n$th coefficient of the Taylor series expansion at
$z=0$ of this function.
\end{proof}

\section{Proofs of the main theorems}
\label{sec:proofs}

We begin by making a number of observations which will be used in the
proofs.  $\psi_t$ and $\phi_t$ each have a single simple pole at
$z=1/\tanh(t)$.  Hence, at any finite time both are analytic in an
open disk containing the closed unit disk and the Taylor coefficients
of their expansions about zero converge to zero exponentially in $n$.

For any fixed $t>0$ the map $\phi_t$ is a fractional linear
transformation which bijectively maps the open unit disk onto itself
and leaves the unit circle invariant.  The points $z=1$ and $z=-1$ are
the two fix points.  In addition, for every fixed $z\in D(1)\backslash
\{1\}$ and fixed neighborhood $\mathcal{N}$ of $-1$, there exists a
time $T(z,\mathcal{N})$ such that $\phi_t(z)\in \mathcal{N}\cap D(1)$
for all $t>T(z, \mathcal{N})$.  The behavior of $G_0\circ \phi_t$ in a
neighborhood of the origin will be important in the analysis which
follows. Observe that $\phi_t(0)=-\tanh(t)$ and for sufficiently small
$r>0$, $\{ \rho e^{i\theta} : \theta \in [0,2\pi], \rho \in[0,r]\}$ is
mapped approximately to $\{ -\tanh(t) + \rho (1-\tanh^2(t)) e^{i
  \theta} : \theta \in [0,2\pi], \rho \in [0,r]\}$ and strictly into
the closed disk
\begin{align*}
  \mathcal{E}_r=\Big\{ -\tanh(t) + \rho (1-\tanh^2(t)) e^{i \theta} : 
  \theta \in
  [0,2\pi], \rho \in [0,r/(1-r)]\Big\}\;.
\end{align*}
For sufficiently small $r>0$, $\mathcal{E}_r$ is strictly contained in the
unit disk for all times $t$. Furthermore, $\mathcal{E}_r$ is bounded away
from the boundary by two lines emanating from $-1$ of the form $\{ -1
+ \rho e^{\pm i\theta} : \rho \geq 0\}$ for some fixed $\theta \in
(0,\pi/2)$.

We recall a basic fact from complex analysis which will be used
repeatedly in the arguments that follow.  To show that the Taylor
coefficients about zero of $G_0 \circ \phi_t$ converges to those of
$F$ as $t \rightarrow \infty$ it is sufficient that for all $t \geq
0$, $F$ and $G_0 \circ \phi_t$ are analytic in a fixed, $t$
independent neighborhood of the origin and that $G_0 \circ \phi_t$
converges uniformly to $F$ on that neighborhood.

\begin{proof}[Theorem \ref{thm:Bexistance}]  
  In order that equation~\eqref{eq:anReconstruction} be well defined,
  $G(z,t)$ needs to be analytic in a neighborhood of $z=0$.  Since
  $\psi_t$ is analytic for all $z\in D(1)$ and each $t >0$, the
  analyticity of $G(z,t)$ about $z=0$ is equivalent to the analyticity
  of $G_0(z)$ about $\phi_t(0)=-\tanh(t)$. As $t$ increases,
  $\phi_t(0)$ decreases monotonically along the real axis from $0$ to
  $-1$.  We need only show that $G_0$ is analytic in an open
  neighborhood of $[-\tanh(t),0]$ in order to complete the proof that
  equation~\eqref{eq:anReconstruction} is well defined for all $t <
  t_*$. $G_0$ is analytic in a neighborhood of the closed interval
  $[-\tanh(t),0]$ since $[-\tanh(t),0] \subset (\alpha,0]$.  Appealing
  to the arguments stated at the beginning of this section we see that
  the image of a small ball about the origin under the mapping
  $\phi_t$ lies in a thin strip about $[-\tanh(t),0]$. Hence, the
  reconstruction formula of equation~\eqref{eq:anReconstruction} is
  well defined because $\Gamma$ can be deformed to lie in a
  sufficiently small ball about the origin.

  If $t_* < \infty$, then $G(z, \tanh( t_*))$ fails to be analytic at
  $z=0$ since $G_0(z)$ in not analytic at $\alpha=-\tanh(t_*)$; we
  cannot continue the solution in this case.

  To see that the $a_n(t)$ (defined as in the statement of the
  theorem) do define a solution, observe that by the definition of
  $G(z,t)$ and integration by parts
  \begin{align*}
    \dot a_n(t) &=  \oint_\Gamma \frac{\d_tG(z,t)}{2\pi i z^{n}} dz
    =  \oint_\Gamma \frac{ (z^2 -1) \d_z G(z,t) + z G(z,t)
    }{2\pi i z^{n}} dz\\
    &=(n-1)\oint_\Gamma \frac{G(z,t)
    }{2\pi i z^{n-1}} dz - n\oint_\Gamma \frac{G(z,t)
    }{2\pi i z^{n+1}} dz= (n-1)a_{n-1}(t) - na_{n+1}(t).
  \end{align*}
  This shows that $a_n(t)$ given by~(\ref{eq:anReconstruction}) is indeed a
  solution of~(\ref{eq:B}) for the initial condition $a_n(0)$  as
  long as $G(z,t)$ is analytic around $z=0$.
  
  Finally, to show that the $a_n(t)$ defined by
  equation~(\ref{eq:anReconstruction}) is the unique solution
  of~(\ref{eq:B}) for the initial condition $a_n(0)$, note that if two
  different solutions exist for the same initial condition, then their
  associated $G(z,t)$ must both satisfy~(\ref{eq:GeqForB}) for the
  same initial condition $G_0(z)$.  Since the solution
  of~(\ref{eq:GeqForB}) is unique, this leads to a contradiction.
\end{proof}

\begin{proof}[Theorem \ref{thm:gozero}]
  Appealing to the discussion at the beginning of the section and the
  fact that $\cosh(t) \sim \frac12 e^t$ as $t\rightarrow \infty$, it
  is enough to show that $\psi_t(z)(G_0 \circ \phi_t)(z)$ converges
  uniformly to $G^+(-1)/(1-z)$ on some neighborhood of the origin.
  First, observe that $\psi_t(z)$ converges to $1/(1-z)$ as $t
  \rightarrow \infty$ uniformly on any disk contained within the unit
  disk.
  
  From the discussion at the beginning of the section we see that for
  all $t>0$ the disk of radius $1/10$ is mapped to a disk contained
  entirely in the open unit disk and bounded away from the unit circle
  by lines emanating from $-1$ of a constant angle. Hence, we have
  \begin{align*}
   \lim_{t \rightarrow \infty} \sup_{|z|<1/10} 
   |G_0(\phi_t(z)) - G_0^+(-1)|=0\;.
  \end{align*}
\end{proof}

\begin{proof}[Theorem \ref{thm:sobbound}]
   Since $(a_1(0),a_2(0),\ldots)$ is square summable,
   $g(r,\theta):=G_0( r e^{i
     \theta})$ is in $L^2(d\theta)$ of the unit circle for all $r \in
   [0,1]$.
Hence, by Plancherel's theorem,
\begin{equation*}
  \sum_{n=1}^\infty |a_n(t)|^2 = \frac{1}{2\pi}\int_0^{2\pi}  |G(
  e^{i\theta},t)|^2 d\theta = \frac{1}{2\pi} \int_0^{2\pi}
  \frac{|\psi_t(e^{i\theta})|^2}{\cosh^2(t)} |G_0(
  \phi_t(e^{i\theta}))|^2 d\theta  
\end{equation*}
Introducing the change of variable $e^{i \eta}=  \phi_t(e^{i\theta})$, we see that
\begin{align*}
  \sum_{n=1}^\infty |a_n(t)|^2 &=\frac{1}{2\pi} \int_0^{2\pi}
  \frac{|\psi_t(\phi_t^{-1}( e^{i\eta}))|^2 }{\cosh^2(t)} |G_0(
  e^{i\eta})|^2 \frac{1}{|\phi_t'( \phi_t^{-1}( e^{i\eta}))|} d\eta\\
  &= \frac1{2\pi}\int_0^{2\pi} |G_0( e^{i\eta})|^2 d\eta =
  \sum_{n=1}^\infty |a_n(0)|^2 \;.
\end{align*}
To obtain the bounds on the weighted norms, we notice that for $s\in
\mathbb{N}$ 
\begin{equation*}
  \sum_{n=1}^\infty n^{2s} |a_n(t)|^2 = \frac1{2\pi}\int_0^{2\pi}  
  \left|\frac{\partial^s
    G}{\partial \theta^s}( e^{i\theta},t)\right|^2 d\theta \;.
\end{equation*}
Using the same change of variable one easily shows that this term can
be bounded in terms of the $L^2$ norms of $\frac{\partial^s \
}{\partial \theta^s} G_0( e^{i\theta})$ for $r\leq s$. By assumption, these
norms are finite.
\end{proof}

\begin{proof}[Theorem \ref{thm:conservesAnalytic}] 
  Fix $T>0$. By the considerations at the beginning of the section,
  one sees that for $t\in [0,T]$ the map $\psi_t(z)(G_0 \circ
  \phi_t)(z)$ remains analytic in $D(1+\eta_1)$ for a sufficiently
  small $\eta_1$ depending only on $T$ and $\eta$. More precisely,
  $\eta_1$ is picked to ensure that $D(1+\eta_1)$ maps into
  $D(1+\eta)$ for all $t\in [0,T]$; the only remaining constraint on
  $\eta_1$ is that $1+\eta_1 < \frac{1}{\tanh{T}}$ so that $\psi_t(z)$
  is also analytic in $D(1+\eta_1)$.  Hence, the power series
  converges absolutely on $D(1+\eta_1/2)$ and $|a_n(t)| \leq
  C(1+\eta_1/2)^n$ for all $n \in \NN$ for some $C>0$.
\end{proof}

\begin{proof}[Theorem \ref{th:2}, Theorem
  \ref{th:remainderOfDisk} and Corollary \ref{cor:1}] Theorem
  \ref{th:2} is a special case of Theorem \ref{th:remainderOfDisk} so
  we concentrate on the later.  By the discussion in the proof of
  Theorem \ref{thm:gozero}, it is clear that for each moment of time
  $t> 0$ there exists a $\eta_1>0$ so that $G(z,t)$ is analytic on
  $\Delta(\zeta_t,\eta_1,\theta_1)\backslash\{\zeta_t\}$. $\eta_1$ may
  be chosen to be sufficiently small in order to avoid other
  singularities of $G_0$ which initially lie outside
  $\Delta(\zeta,\eta,\theta)\backslash\{\zeta\}$ and approach $D(1)$
  under the dynamics of $\phi_t$. A similar consideration needs to be
  taken into account for $\theta$ and may result in an increase of
  $\theta$ to a new $\theta_1$.  As $z \rightarrow \zeta_t$, we see
  that
  \begin{align*}
    G(z,t)
    &\sim\frac{[\cosh(t)-\zeta_t\sinh(t)]^{\alpha-1}}{[\cosh(t)+\zeta
      \sinh(t)]^\alpha} \frac{A}{(\zeta_t -z)^\alpha}\\
    &=\left[\left(\frac{1+\zeta}{2}\right)e^t +
      \left(\frac{1-\zeta}{2}\right)e^{-t} \right]^{1-2\alpha}
    \frac{A}{(\zeta_t -z)^\alpha}\ .
  \end{align*}
  The result on the asymptotics in $n$ then follows from Theorem
  \ref{thm:taub} in the appendix since $G(z,t)$ is analytic on
  $\Delta(\zeta_t,\eta_1,\theta_1)\backslash\{\zeta_t\}$. The result
  for fixed $n$ as $t\rightarrow \infty$ is just a restatement of
  Theorem~\ref{thm:gozero} in this context. The Corollary follows
  directly from the discussion in section~\ref{sec:anomalousdiss}.
\end{proof}

\begin{proof}[Theorem \ref{th:3} and Corollary \ref{cor:2}]
  The proof is similar to that of Theorem \ref{th:2}. Since
  $z=-1$ is a fixed point for $\phi_t$ for all $t >0$, $G(z,t)$ has a
  singularity at $z=-1$ inherited from $G_0(z)$. Since
  the circle is invariant under $\phi_t$, for sufficiently small
  $\eta_1>0$ and $\theta_1$ sufficiently close to $\pi/2$ we have that
  $G(z,t)$ is analytic on $\Delta(-1,\eta_1,\theta_1)$. Direct
  calculation yields:
  \begin{equation*}
    G(z,t)
    \sim\frac1{\cosh(t)+\sinh(t)}\left[\frac{1+\tanh(t)}{1-\tanh(t)}
    \right]^\alpha \frac{A}{(1+z)^\alpha}=\frac{A
      e^{(2\alpha-1)t}}{(1+z)^\alpha} 
    \quad\text{as}\quad z \rightarrow -1\;.  
  \end{equation*}
We obtain
the quoted result by applying Theorem \ref{thm:taub} from the appendix. 

The asymptotics in time follow from the fact that 
  \begin{equation*}
    G(z,t) \sim
    \frac{ Ae^{(2\alpha-1)t}}{1-z} \left(\frac{1-z}{1+z}\right)^\alpha
    \quad\text{as}\quad t \rightarrow \infty 
  \end{equation*}
and direct expression of the right hand side in a Taylor series in $z$.
Corollary  \ref{cor:2} follows  from the discussion
on anomalous dissipation  in section~\ref{sec:anomalousdiss} and the
above results.
\end{proof}

\section{Inviscid limits}\label{sec:invLim}

We return to the analysis of the inviscid limits of
\eqref{eq:B}. Fixing $p\in \NN$ and defining 
\begin{equation*}
  \Lambda_n= \prod_{k=1}^p (n-k)
\end{equation*}
with the convention that $\Lambda_n=1$ if $p=0$, we consider
\begin{equation}\label{eq:dissipation}
  \dot \alpha_{n,\nu} = -2\nu \Lambda_n \alpha_{n,\nu} + \big[ (n-1)\alpha_{n-1,\nu}
  - n \alpha_{n+1,\nu}\big] 
  +\one_{n=1} \dot W(t)\;. 
\end{equation}
As mentioned in Section \ref{sec:B_Example}, it is straightforward to
see that this system converges to a random variable $\alpha^{**}_\nu =
( \alpha^{**}_{1,\nu},\alpha^{**}_{2,\nu}, \cdots)$. In fact, one has
\begin{equation*}
  \EE  \sum_n \Lambda_n|\alpha_{n,\nu}^{**}|^2 = \frac{1}{\nu}\;.
\end{equation*}
Thus, the system does not display anomalous dissipation; the
dissipation which balances the energy injection (due to the forcing) comes
from the term $ -2\nu n^p \alpha_n(t)$.

Setting 
  \begin{equation*}
    \mathcal{G}_\nu(z,t)=\sum_{n=0}^\infty \alpha_{n+1,\nu}(t) z^n,
  \end{equation*}
  one sees that 
  \begin{equation}
    \label{eq:GeqForBdiss}
    \frac{\partial \mathcal{G}_\nu}{\partial t}= (z^2 -1)
    \frac{\partial \mathcal{G}_\nu}{\partial z}
    + z{  \mathcal{G}_\nu} - 2\nu z^p \frac{\partial^p
      \mathcal{G}_\nu}{\partial z^p} + \dot W(t)\; .  
  \end{equation}
  Using the variation of constants formula we obtain
  \begin{align*}
    \mathcal{G}_\nu(\cdot,t) &=  \mathcal{S}_{t,\nu}
    \mathcal{G}_\nu(\cdot,0) + \int_0^t
    (\mathcal{S}_{t-s,\nu}1)\; dW_s.
  \end{align*}
  We will concentrate on the case $p \in \{0,1\}$. By the method
  of characteristics, we find that
\begin{equation*}
  (\mathcal{S}_{t,\nu}f)(z) =
  \begin{cases}\displaystyle
    \frac{e^{-2\nu t}}{\cosh(t)}\psi_t(z) f\big(
    \phi_t(z)\big) & \text{for } p=0\\ \displaystyle
    \frac{1}{\cosh(\kappa t)}\psi_{\kappa t}(z-\nu) f\big(
    \nu + \kappa \phi_{\kappa t}(z-\nu)\big) & \text{for } p=1   
  \end{cases}
\end{equation*}
where $\kappa^2=1+\nu^2$.

It is interesting to contrast the regularizing effect of the different
terms. When $p=0$, $\mathcal{S}_{t,\nu}$ simply dissipates energy at a
faster rate than $S_t$. When $p=1$, $\mathcal{S}_{t,\nu}$ has a
stronger regularizing effect than $\mathcal{S}_{t,\nu}$, in that the
characteristics are attracted to the circle $(\nu -
\kappa)e^{i\theta}$ inside of the unit disk and the singularity of
$\psi_{\kappa t}(z-\nu)$ stays uniformly bounded outside of the unit
disk for all times.  Hence if $f$ has a radius of convergence greater
than $\nu - \kappa \sim -1 + \nu$ then $S_t^{\nu,1}f$ in analytic on a
disk with radius greater than one all times uniformly.

For fixed $t$, $\mathcal{S}_{t,\nu}f$ converges to $S_t$ as
$\nu\rightarrow 0$ uniformly on a neighborhood of the origin. Since
one also has that $S_tf$, $\mathcal{S}_{t,\nu}f$ all go to zero
uniformly on the open disk as $t \rightarrow \infty$ for $f$ bounded
on the unit disk, we have that $\int_{-\infty}^t
(\mathcal{S}_{t-s,\nu})1\; dW_s$ converges to $\int_{-\infty}^t
(S_{t-s}1)\; dW_s$ in mean squared as $\nu \rightarrow 0$. As before
we are primarily interested in these solutions. In this setting they
are given by:
\begin{align}
  \label{eq:stationaryNuZero1}
  \alpha_{n,\nu}^{**}(t)=
  \begin{cases}\displaystyle
\int_{-\infty}^t e^{-2\nu(t-s)}
  \frac{\tanh(t-s)^{n-1}}{\cosh(t-s)} dW(s) & p=0\\\displaystyle
  \int_{-\infty}^t \frac{\kappa}{[\kappa + \nu \tanh(\kappa(t-s))]^{n}}
    \frac{\tanh(\kappa(t-s))^{n-1}}{\cosh(\kappa(t-s))}dW(s)  & p=1
  \end{cases}
\end{align}

\begin{theorem}
\label{thm:viscosityConverge}
  For all $n \in \mathbb{N}$, $t \in \RR$, and $p=0,1$, one has
  \begin{equation*}
    \lim_{\nu \rightarrow 0} \EE \big[ \alpha_{n,\nu}^{**}(t)-
    a^{**}_n(t)  \big]^2 =  0
  \end{equation*}
  Hence  $\alpha_{n,\nu}^{**}(t)$  converges to $a_n^{**}(t)$ as $\nu 
  \rightarrow 0$ for $p=0,1$. Furthermore, one has the estimates given in 
\eqref{eq:Alphap0} and \eqref{eq:Alphap1}.
\end{theorem}
\begin{proof}[Theorem \ref{thm:viscosityConverge} ]
  Fix any $t$. Consider $\alpha_n(t,\nu,0)$ and $a_n(t)$ starting from
  initial condition zero at time $T$ with $T < t$. As $T \rightarrow
  -\infty$, $\alpha_n(t,\nu,0)$ and $a_n(t)$ converge respectively to
  $\alpha_n^{**}(t,\nu,0)$ and $a^{**}_n(t)$.

  By the same argument as Theorem \ref{th:f2}, one see that
  \eqref{eq:stationaryNuZero1} holds.  Subtracting \eqref{eq:stationaryNuZero1}
  from \eqref{eq:limt2}, one obtains
  \begin{align*}
    \EE \big[ \alpha_n^{**}(t,\nu,0) - a^{**}_n(t) \big]^2 &= 2
    \int_0^\infty \big[
    (1+u)^{-\nu} -1  \big]^2 \frac{u^{2(n-1)}}{(u+2)^{2n}} du\\
    &\leq 2 \int_0^\infty \big[(1+u)^{-\nu} - 1 \big]^2
    \frac{1}{(u+1)^{2}}du= \frac{4 \nu^2}{(1+\nu)(1+2\nu)},
  \end{align*}
  which implies that $\alpha_n^{**}(t,\nu,0) \rightarrow a^{**}_n(t)$
  almost surely as $\nu \rightarrow 0$.   The convergence of in the
  other cases is similar. Applying the It\^o isometry to
  \eqref{eq:stationaryNuZero1} proves the quoted value of $\EE
  \big[\alpha_n^{**}(t,\nu,p) \big]^2$ for $p=0$. The other estimates follow
  from
  \begin{equation*}
    \frac{1}{\kappa + \nu} \leq \frac{1}{\kappa + \nu \tanh(t)} \leq
    \frac{1}{\kappa} 
  \end{equation*}
  which holds for $t \geq 0$.
\end{proof}

\begin{remark}
  At first glance, it might seem more natural to consider the system
\begin{equation*}
  \dot{\tilde\alpha}_{n,\nu} = -2 \nu n^p \tilde\alpha_{n,\nu} + \big[
  (n-1)\tilde\alpha_{n-1,\nu} 
  - n \tilde\alpha_{n+1}\big]  
  +\one_{n=1} \dot W(t)\;. 
\end{equation*}
This leads to the following equation for the generating function
$\mathcal{\tilde G}_\nu$:
 \begin{equation*}
   \frac{\partial \mathcal{\tilde G}_\nu}{\partial t}= (z^2 -1)
   \frac{\partial \mathcal{\tilde G}_\nu}{\partial z}
   + z{  \mathcal{\tilde G}_\nu}- 2\nu z^p \frac{\partial^p
     \mathcal{\tilde G}_\nu}{\partial z^p}  -2 \nu \mathcal{D}^p + \dot W(t)\;.
  \end{equation*}
  where $\mathcal{D}^p$ is $p$ applications of the operator defined by
  $(\mathcal{D} f)(z)= \frac{\partial\ }{\partial z}\big(z f(z)\big)$
  and $\mathcal{D}^0$ is the identity operator. Hence, we see that the
  extra dissipative term contains derivative of all orders less then
  or equal to $p$. Not surprisingly, the result is a mixture of the
  behavior of \eqref{eq:dissipation} for all orders less than or equal
  to $p$. In particular, when $p=1$ the asymptotic (in time) behavior
  is given by
  \begin{equation*}
      \tilde \alpha_{n,\nu}^{**}(t)= \int_{-\infty}^t \frac{\kappa e^{-\nu(t-
      s)}}{[\kappa + \nu \tanh(\kappa(t-s))]^{n}}
    \frac{\tanh(\kappa(t-s))^{n-1}}{\cosh(\kappa(t-s))}dW(s)
  \end{equation*}
  and satisfying the following estimate:
  \begin{equation*}
    \frac{ \kappa^2}{(\kappa +\nu)^{2n+2}} \EE[\alpha_{n,{\nu\kappa}/2}^{**}]^2  \leq\EE
    \big[\tilde \alpha_{n,\nu}^{**} \big]^2\leq \frac{1}{\kappa^{2n}}
    \EE[\alpha_{n,{\nu\kappa}/2}^{**}]^2 \;. 
  \end{equation*}
\end{remark}


\section{A second linear shell model}
\label{sec:general}

We begin the analysis of the second model~\eqref{eq:C} by giving
general conditions for the existence of a unique solution of the
initial value problem. The technique is similar to that used in
Section \ref{sec:sol}.

\begin{theorem}
  \label{th:c}
  Let $\{b_n(0)\}$ be such that
  \begin{equation}
    \label{eq:bnoc1}
   \sum_{n=1}^\infty (-1)^n b_n(0) < \infty.
  \end{equation}
  Then the solution of~(\ref{eq:C})
  exists and is unique for all positive times. It can be represented
  as
\begin{equation}
      \label{eq:solH}
      b_n(t) =  \frac{(-1)^{n+1}}{(2n-1)!}
  \frac{\partial^{2n-1}H}{\partial x^{2n-1}}(0,t)
    \end{equation}
    where 
  \begin{equation}
    \label{eq:Hdef}
    H(x,t) = \EE_x H_0\big(X(t)\big) \exp\Big(-\frac12 \int_0^t X^2(s)ds\Big), 
  \end{equation}
  and $X(t)$ satisfies the stochastic differential equation 
  \begin{equation}
    \label{eq:SDE}
    d X(t) = -X^3(t)dt + \frac{1}{\sqrt2} \sqrt{1-X^4(t)} dW(t).
  \end{equation}
  $\EE_x $ denotes the expectation conditional on $X(0)=x\in[-1,1]$ and 
  \begin{equation}
    \label{eq:H0}
    H_0(x) = \sum_{n\in \NN} (-1)^{n+1} b_n(0) x^{2n-1}.
  \end{equation}
\end{theorem}

\begin{remark}
  Alternatively, $H(x,t)$ can be expressed as 
  \begin{equation}
    \label{eq:seriesH}
    H(x,t) =  \sum_{n\in \NN} (-1)^{n+1} b_n(t) x^{2n-1}.
  \end{equation}
where $b_n(t)$ solves (\ref{eq:C}).
\end{remark}

\begin{remark}
  If the sequence $\{b_n(0)\}$ is monotone and converges to zero as $n
  \rightarrow \infty$ then the condition in \eqref{eq:bnoc1} holds.
\end{remark}

The following theorem summarizes the most
interesting properties of solutions of~(\ref{eq:C}).

\begin{theorem}
  \label{th:propgen}
  Suppose that $b_n(0)$ satisfies~(\ref{eq:bnoc1}). Then, for any positive time
  $t>0$, 
  \begin{equation}
    \label{eq:dissbn}
    \lim_{n\to\infty} (2n+1) b_{2n+1}(t) = \bar C_1(t) \text{ and }
    \lim_{n\to\infty} 2n b_{2n}(t) = \bar C_2(t), 
  \end{equation}
  where $\bar C_1(t), \bar C_2(t)\in \RR$, $\bar C_1(t) ,\bar C_2(t)
  \neq 0$ for all but finitely many $t\in[0,\infty)$. In particular,
  there exists a $T>0$ such that for all $t\geq T$, the solution
  of~(\ref{eq:C}) is dissipative and satisfies
  \begin{equation}
    \label{eq:l2norm}
    \sum_{n=1}^\infty b_n^2(t) < \sum_{n=1}^\infty b^2_n(T) < \infty.
  \end{equation}
In fact, $\sum_{n=1}^\infty b_n^2(t)\to 0$ as $t\rightarrow \infty$.
\end{theorem}

\begin{remark}
  The fact that equation~\eqref{eq:C} dissipates energy at finite
  times is implicit in the representation~\eqref{eq:Hdef}. As time
  grows, the factor $\exp(-\frac12 \int_0^t X^2(s)ds)$ converges to
  zero as $\exp(-ct)$ almost surely for some positive deterministic
  $c$. (This follows from the law of large numbers and the verifiable
  assumption that the process is ergodic.)  Hence, $H(x,t)$ converges
  to zero uniformly in $x$ as $t \rightarrow \infty$.
\end{remark}

Writing (\ref{eq:Hdef}) as $H(x,t) = (T_t H_0)(x)$, it is easy to see
that $T_t$ defines a (Feller) semigroup
with generator $L$ defined by
\begin{equation}
  \label{eq:genera}
  \begin{aligned}
    (Lf)(x) & = \frac{(1-x^4)}4\frac{\partial^2 f}{\partial x^2} - x^3
    \frac{\partial f}{\partial x} - \frac{x^2}2 f\\
    &  = \frac{1}{4} \frac{\partial}{\partial x} 
    \Big( (1-x^4)\frac{\partial f}{\partial x}\Big)- \frac{x^2}2 f.
  \end{aligned}
\end{equation}
for $f\in C^2([-1,1])$. In addition, $H(x,t)$ satisfies
\begin{equation}
  \label{eq:Hpde}
  \frac{\partial H}{\partial t} =  \frac{1}{4} \frac{\partial}{\partial x} 
    \Big( (1-x^4)\frac{\partial H}{\partial x}\Big) - \frac{x^2}2 H, 
\end{equation}
with initial condition $H(x,0) = H_0(x)$ for $x\in[-1,1]$. One can
check that the boundaries at $x=\pm1$ are entrance boundaries 
for (\ref{eq:Hpde}) and  $H(x,t)$ satisfies 
\begin{equation}
  \label{eq:entrance}
  \lim_{x\to\pm1} (1-x^4) \frac{\partial H}{\partial x} =0.
\end{equation}

\begin{proof}[Theorem~\ref{th:c}:]
Noting that 
\begin{displaymath}
  \begin{aligned}
    L x^{2n-1}= -n(n+\tfrac12) x^{2n+1} + (n-1)(n-\tfrac12)x^{2n-3},
  \end{aligned}
\end{displaymath}
we compute
\begin{displaymath}
  \begin{aligned}
    &\sum_{n=1}^\infty  (-1)^{n+1} \dot b_n(t) x^{2n-1}\\
    & = \sum_{n=1}^\infty(-1)^{n+1}\left((n-1)(n-\tfrac12) b_{n-1}(t)
      - n(n+\tfrac12) b_{n+1}(t) \right)x^{2n-1}\\
    & =\sum_{n=1}^\infty (-1)^{n+1} b_n(t) \left( - n(n+\tfrac12)
      x^{2n+1}
      + (n-1)(n-\tfrac12) x^{2n-3}\right)\\
    & =\sum_{n=1}^\infty  (-1)^{n+1} b_n(t) L(x^{2n-1}).
  \end{aligned}
\end{displaymath}

\end{proof}

\begin{proof}[Theorem~\ref{th:propgen}:]
Associated with~(\ref{eq:Hpde}) we have the eigenvalue problem
\begin{equation}
  \label{eq:spectrum}
  -\lambda \phi =  \frac{1}{4} \frac{d}{d x} 
    \Big( (1-x^4)\frac{d\phi}{dx}\Big) - \frac{x^2}2 \phi, 
\end{equation}
subject to the boundary conditions
\begin{equation*}
  \lim_{x\to\pm1} (1-x^4) \frac{\partial \phi}{\partial x} =0.
\end{equation*}
It is straightforward to see that the operator in~(\ref{eq:spectrum})
equipped with the boundary condition in~(\ref{eq:entrance}) is
self-adjoint in $L^2[-1,1]$. We now explain why this operator has
discrete spectrum. A standard calculation shows that the boundary is
an ``entrance boundary'' in the sense of Feller
(\cite{Feller60GDO,McKean56ESC}), i.e. the diffusion \eqref{eq:SDE},
if started from the boundary, enters $(-1,1)$ and does not return to
the boundary.

Define $L^V=\frac14 \partial_x (1- x^4) \partial_x + V(x)$ where
$V(x)= -\frac{x^2}2$. By standard PDE theory,
\begin{equation*}
  \partial_t u = L^Vu,
\end{equation*}
subject to the condition $\lim_{t\rightarrow 0 }u(t,x)=\delta_y(x)$,
has smooth solution in $(-1,1)$ for any $t>0$. We denote this
solution by $p_t^V (x,y)$. For a fixed $t>0$, $p_t^V(x,y)$ is Lipschitz
for $x\in [-1+\epsilon, 1- \epsilon]$ and $y\in [-1,1]$ with a fixed
Lipschitz constant $C_t^\epsilon$ and $\sup_{x,y\in [-1,1]}
|p_t^V(x,y)| < D_t$. Consider the solution to the following initial
value problem: Let $f\in L^2([-1,1])$ and solve
\begin{equation*}
  \partial_t u = L^Vu \text{ with } u(0,x)=f(x).
\end{equation*} 
The solution is given by
\begin{equation*}
  u(t,x)=\int p_t^V(x,y)f(y)dy.
\end{equation*}
$u(t,x)$ is Lipschitz for $x\in [-1+\epsilon, 1- \epsilon]$ as the
following simple estimate shows.
\begin{align*}
  |u(t,x)-u(t,x')| &= \left | \int_{-1}^1 (p_t^V(x,y) -p_t^V (x',y))
    f(y)dy \right | \leq \int_{-1}^1 |p_t^V(x,y) - p_t^V(x',y)| |f(y)|
  dy \\ 
  &\leq C_t^\epsilon  |x-x'| \int_{-1}^1 |f(y)| dy \leq 2 C_t^\epsilon
  |x-x'| \left( \int_{-1}^1 |f(y)|^2 dy \right )^\frac12.    
\end{align*}
$u(t,x)$ is also bounded in terms of $\| f\|_2$ as follows:
\begin{equation*} 
  |u(t,x)|= \left | \int_{-1}^1 p_t^V(x,y)
    f(y) dy \right | \leq D_t \int_{-1}^1 |f(y)| dy \leq 2 D_t \left (
    \int_{-1}^1 |f(y)|^2 \right )^\frac12.
\end{equation*} 
As one can see by a Cantor diagonalization argument in intervals $I_n
= [-1+\frac1{n}, 1- \frac1{n}]$, $T_t=e^{t L^V}$ is a compact
self-adjoint operator. Therefore, the spectrum of $L^V$ is discrete.

Note that the lowest eigenvalue has the following variational representation:
\begin{equation*}
  \lambda = \inf_{\phi} 
  \frac{\int_{-1}^1  \big(\frac14(1-x^4) 
    (\phi'(x))^2+\frac12 x^2 \phi^2(x)\big) dx}{
    \int_{-1}^1  \phi^2(x)dx},
\end{equation*}
where the infimum is taken over $L^2$ equipped with the boundary
conditions~\eqref{eq:entrance}. This shows that the spectrum is
strictly positive.  Let $\{\phi_k(x),\lambda_k\}_{k\in\NN}$ be the
pair of eigenfunction and eigenvalues such that each $\phi_k(x)$ is
odd in $x$ (the even ones do not matter since the initial condition
$H_0(x)$ of~(\ref{eq:Hpde}) is odd from~(\ref{eq:H0})). The solution
of~(\ref{eq:Hpde}) can be represented as
\begin{equation}
  \label{eq:solrepphi}
  H(x,t) = \sum_{k\in \NN} h_k e^{-\lambda_kt} \phi_k(x),
\end{equation}
where
\begin{equation*}
  h_k = \int_{-1}^1 H_0(x) \phi_k(x)dx.
\end{equation*}
In turn, (\ref{eq:solrepphi}) implies that
\begin{equation}
  \label{eq:bncnk}
  b_n(t) = \sum_{k\in \NN} h_k e^{-\lambda_kt} p^k_n
\end{equation}
where $p^k_n$ is defined by
\begin{equation}
  \label{eq:pnk}
  \phi_k(x) = \sum_{n=1}^\infty (-1)^{n+1} p^k_n x^{2n-1}.
\end{equation}
The $p^k_n$  satisfy the following recurrence relation inherited
from~(\ref{eq:C}):
\begin{equation}
  \label{eq:recpnk}
  -\lambda_k p^k_n = (n-1)(n-\tfrac12) p^k_{n-1} - n (n+\tfrac12) p^k_{n+1}, 
  \qquad n\in \NN, \ p_0^k =0.
\end{equation}
The following lemma describes the asymptotic behavior of $np_n$.

\begin{lemma}
  For every $\lambda_k>0$, the recurrence relation
  in~(\ref{eq:recpnk}) implies that
  \begin{equation}
    \label{eq:pknasympt}
    \lim_{n\to\infty} (2n+1)p^k_{2n+1} = c^1_k \text{ and }  
    \lim_{n\to\infty} (2n)p^k_{2n} = c^2_k 
  \end{equation}
  where $c^1_k$ and $c^2_k$ are nonzero constant whose sign is the
  same as that of $p^k_1$.
\end{lemma}

\begin{proof}
 Assume $p_1>0$ and  write~(\ref{eq:recpnk}) as
\begin{equation*}
  p^k_{n+1} = \frac{\lambda_k}{n (n+\tfrac12)} p^k_n + 
  \frac{(n-1)(n-\tfrac12)}{n (n+\tfrac12)} p^k_{n-1}
  \qquad n\in \NN, \ p_0^k =0.
\end{equation*}
We omit the index $k$ in this proof as it plays no role. For
sufficiently large $n$ and $C$ depending only on $\lambda$,
\begin{equation*}
  p_{n+1}\leq \left( 1-\frac2{n} + \frac{C}{n^2} \right) \max\{p_n, p_{n-1}\}.
\end{equation*}
This implies $\{p_n\}$ is bounded. More is true:
\begin{equation*}
  p_m \leq \left[ \prod_{l=n}^m \left( 1- \frac2{l} +\frac{C}{l^2} \right) \right]^\frac12  \max\{p_n, p_{n-1}\}.
\end{equation*}
This implies that 
\begin{equation*}
mp_m \leq \exp\left \{ \log m + \frac12 \sum_{l=n}^m \log \left( 1- \frac2{l} + \frac{C}{l^2} \right) \} \max\{ p_n, p_{n-1} \right \},
\end{equation*}
which implies further that $\limsup m p_m < \infty$.
On the other hand,
\begin{equation*}
p_{n+1} \geq \left( 1-\frac2{n+1}      \right) p_{n-1},
\end{equation*}
which implies 
\begin{equation*}
  p_m \geq \left[ \prod_{l=n}^m \left( 1 - \frac2{l+1} \right) \right ]^\frac12  p_n.
\end{equation*}
This implies that 
\begin{equation*}
  mp_m \geq \exp \left \{  \log m  - \frac12  \sum_{l=n}^m \log
    \left(1 - \frac2{l+1} \right) \right \} p_n  \quad\text{if n-m=0
    mod 2}. 
\end{equation*}
Thus, $\liminf mp_m >0$. To show that the the sequences in the theorem
are Cauchy simply compute
\begin{equation*}
  \left| (n+1)p_{n+1} - (n-1)p_{n-1} \right| = \left|
    \frac{\lambda}{n(n+1)} (n+1)p_{n+1}  + (n-1)p_{n-1} \left[
      -\frac12 \frac1{n(n+\frac12)} \right] \right|. 
\end{equation*}
Using the fact that $\{np_n\}$ is bounded in $n$ and summing over $n$
we see that the sequence is Cauchy and have proved the lemma.
\end{proof}

Going back to the proof of Theorem~\ref{th:propgen}, using
(\ref{eq:pknasympt}) in~(\ref{eq:bncnk}) implies (\ref{eq:dissbn}) with
\begin{equation*}
  \bar C_1(t) = \sum_{k\in\NN} e^{-\lambda_kt} h_k c^1_k \text{ and }
  \bar C_2(t) = \sum_{k\in\NN} e^{-\lambda_kt} h_k c^2_k  .
\end{equation*}
Note that there is a $T>0$ so that $\bar C_1(t) \bar C_2(t)>0$ for all
$t\geq T$. Finally, (\ref{eq:dissbn}) implies that
\begin{equation*}
  \frac12 \frac{d}{dt}\sum_{n=1}^\infty b^2_n(t) = 
  -\lim_{N\to\infty} N(N+\tfrac12) b_N(t)b_{N+1}(t) = -\bar C_1(t) \bar C_2(t) <0
\end{equation*}
for all $t\geq T$ which proves~(\ref{eq:l2norm}). 
\end{proof}

We also consider the system of forced equations
\begin{equation}
  \label{eq:Cforce}
  \dot b_n = (n-1)(n-\tfrac12) b_{n-1} - n (n+\tfrac12) b_{n+1} 
  + f(t) \one_{n=m},
\end{equation}
for $n=1,2,\dots$ with boundary condition $b_0(t)=0$ for all $t$ and
 $f(t)$ is either a constant forcing term, $f(t) = 1$, or a
white-noise process, $f(t) = \dot W(t)$.

We have

\begin{theorem}
  \label{th:bb}
  Consider~(\ref{eq:Cforce}) with $f(t)=1$, and initial condition
  $b_n(0)$ satisfying~(\ref{eq:bnoc1}). Then
  \begin{equation*}
    \lim_{t\to\infty} b_n(t) =  b_n^* \equiv 
    \sum_{k\in\NN} \frac{d_k p^k_n}{\lambda_k}
  \end{equation*}
where $p^n_k$ is defined by~(\ref{eq:pnk}) and 
\begin{equation*}
  d_k = \int_{-1}^1 \phi_k(x) x^m dx
\end{equation*}
\end{theorem}
In particular, $b^*_n$ satisfies
\begin{equation*}
  \lim_{n\to\infty} (2n+1) b^*_{2n+1} = C^*_1 >0 \text{ and }
  \lim_{n\to\infty} (2n) b^*_{2n} = C^*_2 >0. 
\end{equation*}

\begin{theorem}
  \label{th:bbb}
  Consider~(\ref{eq:Cforce}) with $f(t)=\dot W(t)$, and initial condition
  $b_n(-T)$ satisfying~(\ref{eq:bnoc1}). Then
  \begin{equation*}
    \lim_{T\to\infty} b_n(t) =  b_n^{**}(t) \equiv 
    \sum_{k\in\NN} d_kp^k_n \int_{-\infty}^t e^{-\lambda_k s} dW(s)
  \end{equation*}
where $p^k_n$ is defined by~(\ref{eq:pnk}).
\end{theorem}

In particular, $b^{**}(t)$ is a Gaussian process with mean zero and covariance
\begin{equation*}
  \EE b^{**}_n(t) b^{**}_m(t) = \sum_{k,k'\in\NN} 
  \frac{d_kd_{k'}p^k_n p^{k'}_m}{\lambda_k+\lambda_k'}
\end{equation*}
and we have
\begin{equation*}
  \lim_{n\to\infty} (2n+1)^2 \EE (b^{**}_{2n+1}(t))^2 = C^{**}_1 >0
  \text{ and } \lim_{n\to\infty} (2n)^2 \EE (b^{**}_{2n}(t))^2 =
  C^{**}_2 >0 . 
\end{equation*}


\appendix

\section{Estimates on Taylor coefficients}
\label{sec:tauberian}

For the reader's convince, we now state a theorem on the asymptotic of
Taylor's series which can be found in \cite{FlajoletOdlyzko90SAG}.
.

\begin{theorem}\label{thm:taub}
  Let $\Delta(\zeta,\eta,\theta)$ be as in
  \eqref{eq:detlaRegion}. Assume that $f(z)$ is analytic in
  $\Delta(\zeta,\eta,\theta)\backslash \{\zeta\}$ for some $\zeta \in \CC$,
  $\eta >0$, and $0<\theta < \pi/2$. If 
  \begin{equation*}
    f(z) \sim \frac{K}{(\zeta -z)^\alpha} \qquad 
    \text{as}\qquad z \rightarrow \zeta
  \end{equation*}
for some $K >0$ and $\alpha \not \in \{0,-1,-2,\cdots\}$ then
\begin{equation*}
  f_n \sim \frac{K}{\Gamma(\alpha)} \frac{n^{\alpha-1}}{\zeta^{n+\alpha}}\
\end{equation*}
where $f_n$ is the $n$-th Taylor coefficient of $f(z)$ about $z=0$.
\end{theorem}

\begin{acknowledge} 
  We thank Percy Deift, Charles Fefferman, Stephanos Venakides and Xin
  Zhou for useful conservations.  J. M. is supported in part by the
  Sloan Foundation and by an NSF CAREER award. T. S is supported in
  part by NSF grant DMS05-53403. E.  V.-E.  is supported in part by
  NSF grants DMS02-09959 and DMS02-39625, and by ONR grant
  N00014-04-1-0565.
\end{acknowledge}


\begin{thebibliography}{FGV01}

\bibitem[CET94]{ConstantinETiti94OCE}
Peter Constantin, Weinan E, and Edriss~S. Titi.
\newblock Onsager's conjecture on the energy conservation for solutions of
  {E}uler's equation.
\newblock {\em Comm. Math. Phys.}, 165(1):207--209, 1994.

\bibitem[DR00]{DuchonRobert00IED}
Jean Duchon and Raoul Robert.
\newblock Inertial energy dissipation for weak solutions of incompressible
  {E}uler and {N}avier-{S}tokes equations.
\newblock {\em Nonlinearity}, 13(1):249--255, 2000.

\bibitem[E01]{E01SH} Weinan E.  \newblock Stochastic hydrodynamics.
  \newblock In {\em Current developments in mathematics, 2000}, pages
  109--147.  Int. Press, Somerville, MA, 2001.

\bibitem[Eyi01]{Eyink01DTE}
Gregory~L. Eyink.
\newblock Dissipation in turbulent solutions of 2{D} {E}uler equations.
\newblock {\em Nonlinearity}, 14(4):787--802, 2001.

\bibitem[Fel54]{Feller60GDO}
William Feller.
\newblock The general diffusion operator and positivity preserving semi-groups
  in one dimension.
\newblock {\em Ann. of Math. (2)}, 60:417--436, 1954.

\bibitem[FGV01]{FalkovichGaweddzkiVergassola01PFF}
G.~Falkovich, K.~Gaw{\c{e}}dzki, and M.~Vergassola.
\newblock Particles and fields in fluid turbulence.
\newblock {\em Rev. Modern Phys.}, 73(4):913--975, 2001.

\bibitem[FO90]{FlajoletOdlyzko90SAG}
Philippe Flajolet and Andrew Odlyzko.
\newblock Singularity analysis of generating functions.
\newblock {\em SIAM J. Discrete Math.}, 3(2):216--240, 1990.

\bibitem[Fri95]{Frisch95}
Uriel Frisch.
\newblock {\em Turbulence}.
\newblock Cambridge University Press, Cambridge, 1995.
\newblock The legacy of A. N. Kolmogorov.

\bibitem[Hil01]{Hilberdink01TPS}
Titus Hilberdink.
\newblock A {T}auberian theorem for power series.
\newblock {\em Arch. Math. (Basel)}, 77(4):354--359, 2001.

\bibitem[McK56]{McKean56ESC}
Henry~P. McKean, Jr.
\newblock Elementary solutions for certain parabolic partial differential
  equations.
\newblock {\em Trans. Amer. Math. Soc.}, 82:519--548, 1956.

\bibitem[Sri05]{woodshole}
Ravi Srinivasan.
\newblock Simple models with cascade of energy and anomalous dissipation.
\newblock In Oliver Buhler and Charles Doering, editors, {\em Fast times and
  fine scales}, Woods Hole Oceanographic Institution Technical Reports. Woods
  Hole Oceanographic Institution, 2005.

\end{thebibliography}
\end{document}